%% file: main.tex
\PassOptionsToPackage{table}{xcolor}

\documentclass[preprint,12pt,authoryear]{elsarticle}

\input{setup}
\input{commands}

\journal{Journal of Systems and Software}

\begin{document}

\begin{frontmatter}

\title{A Systematic Literature Review on the Impact of Formatting Elements on Code Legibility}

\author[ufpe,ifpe]{Delano Oliveira\corref{contrib}}
\ead{dho@cin.ufpe.br}
\author[ufpe]{Reydne Santos\corref{contrib}}
\ead{reydne.bruno@gmail.com}
\author[vua]{Fernanda Madeiral}
\ead{fer.madeiral@gmail.com}
\author[tokyo]{Hidehiko Masuhara}
\ead{masuhara@acm.org}
\author[utrecht,ufpe]{Fernando Castor}
\ead{f.j.castordelimafilho@uu.nl}

\cortext[contrib]{Authors contributed equally.}

\affiliation[ufpe]{organization={Federal University of Pernambuco},
            city={Recife},
            country={Brazil}}
\affiliation[ifpe]{organization={Federal Institute of Pernambuco},
            city={Recife},
            country={Brazil}}
\affiliation[vua]{organization={Vrije Universiteit Amsterdam},
            city={Amsterdam},
            country={Netherlands}}
\affiliation[tokyo]{organization={Tokyo Institute of Technology},
            city={Tokyo},
            country={Japan}}
\affiliation[utrecht]{organization={Utrecht University},
            city={Utrecht},
            country={Netherlands}}

\begin{abstract}
\textbf{Context:} Software programs can be written in different but functionally equivalent ways. Even though previous research has compared specific formatting elements to find out which alternatives affect code legibility, seeing the bigger picture of what makes code more or less legible is challenging.
\textbf{Goal:} We aim to find which formatting elements have been investigated in empirical studies and which alternatives were found to be more legible for human subjects.
\textbf{Method:} We conducted a systematic literature review and identified \nbPapersFinalSelectionAndSnowball papers containing human-centric studies that directly compared alternative formatting elements. We analyzed and organized these formatting elements using a card-sorting method.
\textbf{Results:} We identified \nbFactors formatting elements (e.g., indentation) and \nbLevels levels of formatting elements (e.g., two-space indentation), which are about formatting styles, spacing, block delimiters, long or complex code lines, and word boundary styles. While some levels were found to be statistically better than other equivalent ones in terms of code legibility, e.g., appropriate use of indentation with blocks, others were not, e.g., formatting layout. For identifier style, we found divergent results, where one study found a significant difference in favor of camel case, while another study found a positive result in favor of snake case.
\textbf{Conclusion:} The number of identified papers, some of which are outdated, and the many null and contradictory results emphasize the relative lack of work in this area and underline the importance of more research. There is much to be understood about how formatting elements influence code legibility before the creation of guidelines and automated aids to help developers make their code more legible.
\end{abstract}



\begin{keyword}
Program understandability \sep Code legibility \sep Formatting elements.

\end{keyword}

\end{frontmatter}


\section{Introduction}
\label{sec:introduction}

Program comprehension is a required activity for any software maintenance and evolution task. Even when documentation is available, software developers have to understand fine-grained elements of the source code to be able to modify it. These elements relate to visual, structural, and semantic characteristics of the source code of a program and may hinder program comprehension. 
 
\textit{Formatting elements}, such as spacing, are \textit{factors} that impact the \textit{legibility} of the source code and, consequently, may affect the ability of developers to identify the elements of the code while reading it. Structural and semantic characteristics, such as programming constructs, impact the \textit{readability} of the source code and may affect the ability of developers to understand it while reading the code. We use the term \textit{understandability} to refer to the ease with which developers are able to extract information from a program that is useful for a software development- or maintenance-related task just by reading its source code. Understandability is impacted by legibility and readability. 

Ensuring the understandability of software projects is a challenging task. Some software organizations (e.g., Google\footnote{\url{https://google.github.io/styleguide/}, last access May 22, 2023.} and Sun Microsystems\footnote{\url{https://www.oracle.com/java/technologies/javase/codeconventions-contents.html}, last access May 22, 2023.}) have adopted code conventions to enforce the usage of software development best practices by their developers. \cite{smit2011} explain that \textit{``code conventions are a body of advice on lexical and syntactic aspects of code, aiming to standardize low-level code design under the assumption that such a systematic approach will make code easier to read, understand, and maintain''}.

There exist guidelines and coding standards for different programming languages, such as the Google Java Style Guide\footnote{\url{https://google.github.io/styleguide/javaguide.html}, last access May 22, 2023.} for the Java language, which describe good practices and well-accepted conventions on how to write code. However, these guides are built based on the intuition and experience of the developers that elaborate them. In fact, \textit{what makes the code more legible or readable is an open question}. Researchers have conducted empirical studies to compare different but functionally equivalent ways of writing code in terms of their legibility and readability. For instance, \cite{Miara1983} compared different levels of indentation (zero, two, four, and six spaces), and \cite{Binkley2013} investigated the usage of camel case vs. underscore for identifier names. These studies provide insight into the influence of different formatting elements on code legibility, but they are parts of a bigger whole that is still unclear.

In this paper, we aim to provide a more comprehensive view of the existing knowledge about the impact of different formatting elements on code legibility.
To do so, we examined empirical studies performed by researchers with human subjects aiming to find which formatting alternatives are the best for the \textbf{legibility} of source code. We conducted a systematic literature review that, from an initial set of \nbPapersFromAllEnginesAndSnowball documents, analyzes \nbPapersFinalSelectionAndSnowball papers. We selected papers that directly compared alternative levels of formatting elements and organized them through a card-sorting process. Finally, for each formatting element, we examined the findings reported by the primary studies considering two aspects of human-centric studies that evaluate code legibility: the activities performed by human subjects (which relate to the cognitive skills required from them) and the response variables employed by the researchers to collect data from the studies~\citep{oliveira2020}.

We identified \nbFactors formatting elements (e.g., indentation) and \nbLevels alternative levels of these factors (e.g., two-space indentation) in \nbComparisons comparisons (e.g., one comparison considered zero, two, four, and six indentation levels). These formatting elements are about formatting styles (e.g., formatting layout), spacing (e.g., indentation), block delimiters (e.g., block delimiter visibility), long or complex code line (e.g., statements per line), and word boundary styles (i.e., identifier styles).

All comparisons for \numberToBeChecked{four} factors showed statistically significant results in favor of the same level (e.g., book format for formatting style). For other \numberToBeChecked{four} factors, there were comparisons where statistically significant differences between the levels were found, while other comparisons did not show differences. For instance, in comparisons involving horizontal spacing, \cite{Miara1983} found out that two-space indentation leads to code that is easier to understand than zero, four, or six spaces. However, recent works~\citep{Santos2018,Bauer2019} did not find differences between indentation levels, though the studied source code snippets of these studies were from different programming languages. For \numberToBeChecked{the identifier style factor}, there were divergent results. \cite{Sharif2010} found that snake case is the best alternative for identifier names. However, \cite{Binkley2013} found out that the usage of camel case is better than snake case in one study, and no statistical difference between the two formatting alternatives was found in a second study. For the remaining \numberToBeChecked{four} factors, e.g., blank space around operators and parameters~\citep{sampaio2016}, no statistically significant differences were found.

This study highlights that our current understanding of code legibility is limited. The area is \textbf{immature}, i.e., there are few studies on the topic, the studies that do exist are \textbf{narrow in scope}, i.e., they employ a restricted and small set of approaches to evaluate alternative solutions, and the results of many studies are \textbf{inconclusive}, in part due to lack of statistical power. Furthermore, \nbOldPapers/\nbPapersFinalSelectionAndSnowball papers we analyzed are \textbf{outdated}, having been published 30 or more years ago. This indicates the need for more research in this area.
Ultimately, understating the impact of formatting elements on code legibility would allow the creation of guidelines and automated aids, e.g., linters and recommendation systems, to help developers during programming activities and make their code more legible.

\section{Background} \label{sec:background}

In a previous work~\citep{oliveira2020}, we conducted a systematic literature review to define how researchers measure code legibility and readability. To achieve that, we examined primary studies where human subjects are asked to perform programming-related tasks involving comparisons between alternative programming constructs, idioms, and styles. These alternatives are different but functionally equivalent ways of writing code, e.g., recursive vs. iterative code~\citep{Benander1996} and abbreviated vs. word identifier names~\citep{Hofmeister2019}. The goal of those comparisons is to find the most legible or readable ways of writing code. We did not, however, investigate the alternatives themselves, which is the goal of the current work.

In this section, we summarize the main three elements of that paper that we build upon. First, we introduce the concepts of code legibility and readability in~\autoref{sec:legibility_and_readability}. We discuss how Software Engineering researchers define these two terms, and we bring insights from other areas of knowledge to make a clear distinction between them. In~\autoref{sec:tasks_response_variables}, we provide background on the tasks performed by humans and response variables collected in studies that compared functionally-equivalent source code alternatives, as found by \cite{oliveira2020}. The combination of these tasks and response variables builds a landscape of the different ways in which researchers define and measure code legibility and readability in existing studies. Finally, in~\autoref{sec:learning_activity}, we present how we leverage a learning taxonomy~\citep{bloom1956taxonomy,fuller2007developing} to model program comprehension tasks so that researchers can clearly express the cognitive skills they intend to evaluate with a study. More specifically, we review a set of activities that decompose the tasks performed by humans in the primary studies of our previous study \citep{oliveira2020}.
 
\subsection{Legibility and Readability}\label{sec:legibility_and_readability}

In software engineering, the terms readability and legibility have overlapping meanings. For example, \cite{Buse2010} define ``\textit{\textbf{readability} as a human judgment of how easy a text is to understand}''. In a similar vein, \cite{de2003best} affirm that ``\textit{\textbf{legibility} is fundamental to code maintenance; if source code is written in a complex way, understanding it will require much more effort}''.

In linguistics, the concept of text comprehension is similar to program comprehension in software engineering. \cite{gough1986decoding} state that ``\textit{comprehension (not reading comprehension, but rather linguistic comprehension) is the process by which given lexical (i.e., word) information, sentences and discourses are interpreted}''. However, \cite{hoover1990simple} further elaborate on that definition and claim that ``\textit{decoding and linguistic comprehension are separate components of reading skill}''. This claim highlights the existence of two separate processes during text comprehension: (i) decoding the words/symbols and (ii) interpreting them and the sentences formed by them. 

\cite{dubay2004principles} separates these two processes and defines them as \textbf{legibility}, which concerns typeface, layout, and other aspects related to the identification of elements in text, and \textbf{readability}, that is, what makes some texts easier to read than others. In a similar vein, for \cite{tekfi1987readability}, legibility studies are mainly concerned with typographic and layout factors, while readability studies concentrate on linguistic factors. 

These two perspectives also apply to programs. We can find both the visual characteristics and linguistic factors in source code, although with inconsistent terminology. 
For example, \cite{daka2015} state that ``\textit{the visual appearance of code in general is referred to as its readability}''. The authors clearly refer to legibility (in the design/linguistics sense) but employ the term ``readability'' possibly because it is more often used in the Software Engineering literature.

Based on the differences between the terms \textit{``readability''} and \textit{``legibility''} that are well-established in other areas such as linguistics~\citep{dubay2004principles}, design~\citep{strizver2013type}, human-computer interaction~\citep{zuffi2007human}, and education~\citep{tekfi1987readability}, we believe that the two terms should have clear, distinct, albeit related, meanings also in the area of Software Engineering. On the one hand, the structural and semantic characteristics of the source code of a program that affect the ability of developers to understand it while reading the code, e.g., programming constructs, coding idioms, and meaningful identifiers, impact its \textbf{readability}. On the other hand, the visual characteristics of the source code of a program,  which affect the ability of developers to identify the elements of the code while reading it, such as line breaks, spacing, alignment, indentation, blank lines, identifier capitalization, impact its \textbf{legibility}. Hereafter, we employ these two terms according to these informal definitions. In this work, we focus on code legibility.

\subsection{Tasks and Response Variables Employed in Human-centric Studies on Code Legibility and Readability}\label{sec:tasks_response_variables}

In our previous study~\citep{oliveira2020}, we examined how researchers have investigated code legibility and readability by asking human subjects to perform tasks on source code and assessing their understanding or effort with response variables. A recent study by~\cite{feitelson2022} also presented the tasks and metrics adopted in studies about code comprehension. Different from our previous study, their focus was narrower: they performed an in-depth analysis of the factors involved in only one of the levels of learning skills (``understanding'') presented by \cite{oliveira2020}, and on concrete activities performed by developers. In this section, we review the tasks and response variables we found during our previous study, which lay a foundation for the study presented in this paper.

\vspace{5pt}
\noindent\textbf{Tasks.}
The essential code comprehension task is code reading. By construction, human-centric studies about code comprehension have at least one reading task where the subject is required to read a code snippet, a set of snippets, or even large, complete programs. In addition, the subjects are also expected to comprehend the code. However, there are different (kinds of) tasks that help researchers measure subject comprehension performance.

A large portion of the studies required subjects to \textbf{provide information about the code}.
For example, \cite{Benander1996} asked the subjects to explain using free-form text what a code snippet does right after having read it.
In some studies, the subjects were asked to answer questions about code characteristics. For example, \cite{Gopstein2017} and \cite{Ajami2019} presented subjects with multiple code snippets and asked them to guess the outputs of these snippets.

Also, some studies required the subjects to \textbf{act on the code}.
Among these studies, subjects were asked to find and fix bugs in the code. For example, \cite{Scanniello2013} asked subjects to do so in programs with different identifier styles.
In other studies, the subjects were asked to modify the code of a working program, i.e., without the need to fix bugs. For example, \cite{Jbara2014b} asked subjects to implement a new feature in a program seen in a previous task.

Lastly, subjects were also asked to give their \textbf{personal opinion}.
In some studies, the subjects were inquired about their personal preferences or gut feeling without any additional task. For example, \cite{Buse2010} asked them to rate (from 1 to 5) how legible or readable a code snippet is.
In other studies, the subjects were asked about their personal opinions while performing other tasks. For example, \cite{ONeal1994} first asked subjects to read a code snippet, then to state if they understood the snippet and provide a description of its functionality.

\vspace{5pt}
\noindent\textbf{Response variables.}
After human subjects perform tasks, researchers employ response variables to measure the subjects' performance.
Depending on the study's goals, methodology, and subjects, response variables vary.
The performance of subjects is measured in some studies in terms of whether they provide correct answers given a task. The \textbf{correctness} of answers may pertain to code structure, semantics, use of algorithms, or program behavior. For example, \cite{Bauer2019} measured the subjects' code understanding by asking them to fill in a questionnaire with multiple-choice questions referring to program output.

Studies that evaluate which code alternative is easier to read often collect the subjects' personal opinions. These response variables are grouped in a category called \textbf{opinion}. For instance, \cite{Santos2018} presented pairs of functionally equivalent snippets to the subjects and asked them to choose which one they think is more readable or legible.

Some researchers measure the \textbf{time} subjects spend performing tasks. There is variety in the way the studies measure time. For \cite{Ajami2019}, \textit{``time is measured from displaying the code until the subject presses the button to indicate he is done''}. \cite{Hofmeister2019} computed the time subjects spent looking at specific parts of a program.

More recent studies collect information about the process of performing the task instead of its outcomes. Multiple studies employ special equipment to track what the subjects see and to monitor the subjects' brain activities during the tasks. The response variables gathered during these processes are \textbf{visual metrics} and \textbf{brain metrics}. For example, \cite{Binkley2013} computed the visual attention, measured as the amount of time during which a subject is looking at a particular area of the screen. \cite{Siegmund2017} used functional magnetic resonance imaging (fMRI) to measure brain activity by detecting changes associated with blood flow.

\subsection{Program Comprehension as a Learning Activity}\label{sec:learning_activity}

The empirical studies analyzed in our previous work \citep{oliveira2020} involve a wide range of tasks to be performed by their subjects, as reviewed in the previous section. All these tasks are conducted to evaluate readability and legibility. However, they demand different cognitive skills from the subjects and, as a consequence, evaluate different aspects of readability and legibility. In our previous work, we attempted to shed light on this topic by analyzing the cognitive skill requirements associated with each kind of task.

According to the Merriam-Webster Thesaurus, to learn something is \textit{``to gain an understanding of''} it. Following along these lines, we treated the problem of program comprehension (or understanding) as a learning problem. We proposed an adaptation of the learning taxonomy devised by \linebreak \cite{fuller2007developing} to the context of program comprehension. \cite{fuller2007developing}'s taxonomy is composed of a set of activities that build upon Bloom's revised taxonomy~\citep{bloomrevised2001} (for educational objectives) and a model that emphasizes that some activities in software development involve acting on knowledge, instead of just learning. Our central idea was to leverage the elements defined by the taxonomy of~\cite{fuller2007developing}, with some adaptions, to identify and name the cognitive skills required by different tasks employed by code readability and legibility studies. The resulting activities are described with examples in \autoref{tab:learning-activities}. We introduced two activities (marked with~``*'') that stem directly from tasks performed by subjects in some of the primary studies we analyzed and required skills that are not covered by the original set of activities.

\input{tables/learning-activities}

\input{tables/tasks-activities-mapping}

We analyzed the tasks that subjects performed in the primary studies and identified which activities from~\cite{fuller2007developing} they require.
\autoref{tab:tasks-activities-mapping} presents the mapping of tasks (rows) to the learning activities (columns).
Moreover, besides the activities in \autoref{tab:learning-activities}, we also considered Giving an Opinion (hereafter, ``Opinion''), which is not part of the taxonomy because it is not a learning activity. The tasks listed in the table were extracted from the previously analyzed studies. Furthermore, the activities to which they are mapped are the ones that were required from subjects in those studies, i.e., there may be other activities associated with those tasks that have not been investigated.

\autoref{tab:tasks-activities-mapping} shows that there is a direct correspondence between some tasks and activities. For example, all the instances of the \textit{``Find and fix bugs in the code''} task involve the Debug activity, and all the tasks that require subjects to provide an opinion are connected to the Opinion activity. In addition, some tasks may be connected to various activities. For instance, \textit{``Modify the code''} may require subjects to Implement, Trace, Inspect, or Adapt the code to be modified. This makes sense: to modify a program, one may have to understand its static elements and its behavior, as well as adapt code elements to be reused. Another example is \textit{``Answer questions about code''}, which often requires subjects to Trace and Inspect code. Furthermore, \textit{``Explain what the code does''} is usually related to Present. Notwithstanding, in some studies~\citep{ONeal1994, Blinman2005}, subjects were presented with multiple descriptions for the same code snippet and asked which one is the most appropriate. This requires the subject to Relate different programs and descriptions. Finally, there are some non-intuitive relationships between tasks and activities. \cite{Chaudhary1980} asked the subjects to reconstruct a program after spending some time looking at it. This is a \textit{``Remember the code''} task where the subjects have to explain what the program does by writing a similar program. It is a mix of Present and Implement. 

\section{Methodology}
\label{sec:methodology}

In this paper, we aim to examine what formatting elements have been investigated and which ones were found to be more legible in human-centric studies. We focus on studies that directly compare two or more functionally equivalent alternative ways of writing code. We address the following research questions in this paper:

\newcommand\rqone{What formatting elements at the source code level have been investigated in human-centric studies?}

\newcommand\rqtwo{Which levels of formatting elements have been found to make the source code more legible?}

\begin{description}
    \item[RQ1] \rqone
    \item[RQ2] \rqtwo
\end{description}

\begin{figure}[t]
    \centering
    \includegraphics[scale=0.68]{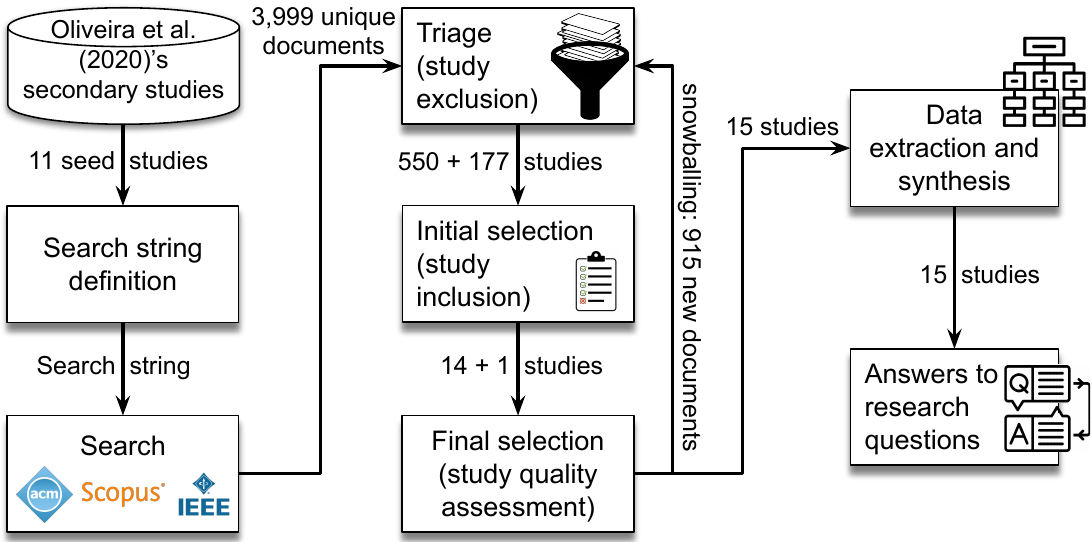}
    \caption{Systematic literature review roadmap.}
    \label{fig:review_roadmap}
\end{figure}

To answer our research questions, we conducted a systematic literature review designed following the guidelines proposed by \cite{Kitchenham2015}. \autoref{fig:review_roadmap} presents the roadmap of our review, including all steps we followed.
First, we performed the selection of studies. We started with a set of papers (\numberToBeChecked{\nbSeedPapers}) about code legibility found in our previous work~\citep{oliveira2020}, which we used as seed papers so that a search string could be defined, and automatic search could be performed on search engines (\autoref{sec:search}). We retrieved \nbPapersFromAllEngines unique documents with the automatic search, which passed through a triage for study exclusion (\autoref{sec:triage}), an initial study selection where inclusion criteria were applied (\autoref{sec:inclusion}), and a final study selection where we evaluated the quality of the studies based on several criteria (\autoref{sec:quality-assessment}). After this process, \nbPapersFinalSelection papers were selected. Then, we performed backward and forward snowballing (\autoref{sec:snowballing}), where we found \nbPapersFromSnowballing new documents, which passed through all the selection steps. We found \nbPapersSnowballingAccepted additional paper, totaling \nbPapersFinalSelectionAndSnowball papers for our review. Finally, the selected \nbPapersFinalSelectionAndSnowball papers were analyzed (\autoref{sec:data_analysis}) for data extraction and synthesis to answer our research questions. We detail these steps in the following sections. 

\vspace{5pt}
\noindent\textbf{Full disclosure on novelty.} Our methodology resembles the one employed in our previous work~\citep{oliveira2020}. However, the search for papers is different, which means the whole process started based on a new set of papers for this work. Moreover, the data analysis and synthesis methods are completely new because we answer different research questions in this paper. A detailed elaboration on the differences between the two works is presented in \autoref{sec:diffs}.

\subsection{Search Strategy}\label{sec:search}

Our search strategy comprises three parts: the selection of papers to be used as seed studies, the definition of a generic search string, and the automatic search in search engines.
First, we selected relevant papers as seeds to define a generic search string with terms that would return relevant papers for answering our research questions. We selected \nbSeedPapers papers from our previous study~\citep{oliveira2020} to be considered as seeds in this one. These papers contain studies that compare two or more ways of writing code by only changing formatting elements to help code understanding in terms of legibility. This is the kind of study we are searching for in this work and, therefore, these papers form the ideal seed set for systematically searching other papers.

We analyzed the title and keywords of the seed papers and extracted the general terms related to our research questions. 
We used the resulting terms to build the following search string:

\begin{center}
\footnotesize
Title(ANY($terms$))~OR~Keywords(ANY($terms$)),

where
$terms = \{$
``code comprehension'',
``code understandability'',
``code understanding'',
``code readability'',
``code complexity'',
``code misunderstanding'',
``code patterns'',
``program comprehension'',
``program understandability'',
``program understanding'',
``program readability'',
``program idioms'',
``program style'',
``program patterns'',
``programmer experience''
$\}$
\end{center}

We did not include terms with ``legibility'' in the search string. Most of the papers with this word in the title or keywords are related to linguistics or computational linguistics. In these fields, researchers use this term with a different meaning than what would be expected in a software engineering paper. Using it in our search string would drastically increase the number of false positives. 

Finally, we performed an automatic search for studies using our generic search string adapted for three search engines: ACM Digital Library\footnote{\url{http://dl.acm.org/}, last access May 22, 2023.}, IEEE Explore\footnote{\url{http://ieeexplore.ieee.org/}, last access May 22, 2023.}, and Scopus\footnote{\url{http://www.scopus.com/}, last access May 22, 2023.}. We retrieved \nbPapersACM, \nbPapersIEEE, and \nbPapersScopus documents, respectively, on \numberToBeChecked{October 10, 2022}. Since a given document might be retrieved from more than one engine, we unified the output of the engines to eliminate duplicates, which resulted in \nbPapersFromAllEngines unique documents. The \nbSeedPapers seed papers were returned by the automatic search.

\subsection{Triage (Study Exclusion)}\label{sec:triage}

The \nbPapersFromAllEngines documents retrieved with our automatic search passed through a triage process so that we could discard clearly irrelevant documents. For that process, we first defined five exclusion criteria:

\begin{description}
    \item[\textbf{EC1}.] The study is clearly irrelevant to our research questions, i.e., it is not primarily related to source code comprehension, readability, legibility, or hard-to-understand code, or does not involve any comparison of different ways of writing code.
    \item[\textbf{EC2}.] The study is not a full paper (e.g., PhD theses or 4-page papers), is not written in English, or was not peer-reviewed: not considering these types of documents in systematic reviews is a common practice \citep{Kitchenham2015}.
    \item[\textbf{EC3}.] The study is about legibility, readability, or understandability metrics.
    \item[\textbf{EC4}.] The study is about program comprehension aids, such as visualizations or other forms of analysis or sensory aids, e.g., trace-based execution, code summarization, and character encodings supporting colors.
    \item[\textbf{EC5}.] The study focuses on accessibility, e.g., targets individuals with visual impairments or neurodiverse developers.
\end{description}

Then, each of the \nbPapersFromAllEngines documents was analyzed by an author of this paper, who checked the title and abstract of the document, and in some cases, the methodology, against the exclusion criteria. 
The documents that did not meet any of the exclusion criteria were directly accepted to enter the next step. The documents that meet at least one exclusion criterion passed through a second round in the triage process, where each document was analyzed by a different author.
At the end of the two rounds, we discarded all documents that were annotated with at least one exclusion criterion in both rounds. We followed this two-round process to mitigate the threat of discarding potentially relevant studies in the triage.
We ended up with \nbPapersAfterExclusionCriteria papers.

\subsection{Initial Selection (Study Inclusion)}\label{sec:inclusion}

After discarding clearly irrelevant documents in the triage process, we applied the following inclusion criteria to the \nbPapersAfterExclusionCriteria papers to build our initial set of papers:

\begin{description}
    \item[\textbf{IC1} (Scope).] The study must be primarily related to \textit{legibility}.
    \item[\textbf{IC2} (Methodology).] The study must be or contain a controlled experiment, quasi-experiment, or survey \textit{involving human subjects}.
    \item [\textbf{IC3} (Comparison).] The study must directly \textit{compare alternative formatting elements in terms of code legibility}, and the alternatives must be clearly identifiable.
    \item [\textbf{IC4} (Granularity).] The study must target fine-grained program elements and low-level/limited-scope programming activities. Not design or comments, but implementation.
\end{description}

The application of the inclusion criteria to a paper often requires reading not only the title and abstract as in the triage process, but also sections of introduction, methodology, and conclusion. If a given paper violates at least one inclusion criterion, the paper is annotated with ``not acceptable''. When there are doubts about the inclusion of a paper, the paper is annotated with ``maybe'' for further discussion. We also performed this step in two rounds, but differently from the triage, all papers in this step were independently analyzed by two different authors.
At the end of this step, the papers annotated with ``acceptable'' in both rounds were directly selected, and papers annotated with ``not acceptable'' in both rounds were rejected. All the other cases were discussed by all authors in live sessions to reach consensus. We ended up with \nbPapersAfterInclusionCriteria papers.

\vspace{5pt}
\noindent\textit{Deprecated studies.}
We identified some papers that we refer to as \textit{deprecated}. A paper is deprecated if it is extended by another paper that we selected. For instance, the paper of~\cite{binkley2009b} was extended in a subsequent paper \citep{Binkley2013}. In this case, we consider the former to be deprecated and only take the latter into account.

\subsection{Study Quality Assessment}\label{sec:quality-assessment}

The selected \nbPapersAfterInclusionCriteria papers passed through a final selection step, aiming to identify low-quality papers for removal.
To do so, we elaborated nine questions that were answered for each paper. We adapted these questions from the work of~\cite{keele2007guidelines}.
Three questions were about study design, e.g., \textit{``are the aims clearly stated?''}, four questions were about analysis, e.g., \textit{``are the data collection methods adequately described?''}, and two questions were about rigor, e.g., \textit{``do the researchers explain the threats to the study validity?''}. Each question was answered with one of the following possible answers: yes (1), partially (0.5), and no (0). The sum of the answers for a given paper is its quality score. The maximum is, therefore, 9. If a paper scores 0.5 or better in all questions, its overall score is 4.5 or more. Thus, we defined that a paper should score at least 4.5 to be kept in our list of papers.

Each paper was assessed by one of the authors of this paper. At the beginning of this step, each author selected one paper, performed the quality assessment, and justified to the other authors the given score for each question in a live discussion. This procedure allows us to align our understanding of the questions and avoid misleading assessments. The scores of the papers were: \numberToBeChecked{$min = 5$, $median = 8$, $max = 9$}. Since the minimum score for keeping a paper is 4.5 and no paper scored less than 5, no papers were removed because of low quality.

\subsection{Snowballing}\label{sec:snowballing}

To make sure our study is comprehensive, we carried out a snowballing process. To do so, we collected new papers by gathering all the references used in the included papers (backward snowballing) and all the papers that cited the included papers in our review (forward snowballing). We used Google Scholar to extract the citations of included papers. The newly collected papers passed through the triage (study exclusion), initial selection (study inclusion), and final selection (study quality assessment). At this stage, only papers published from the 1990s onwards were included because that period marked a significant advancement in the field of software engineering with the emergence of object-oriented programming. We repeated this process until no new paper was found. We performed \nbSnowballingLevels complete iteration, and no new paper was found for a second one. With the initial set of \nbPapersFinalSelection selected papers, we gathered \nbPapersFromSnowballing new documents. After analyzing them by considering the exclusion and inclusion criteria, we found \nbPapersSnowballingAccepted new paper, totaling \nbPapersFinalSelectionAndSnowball papers for our review.

\subsection{Data Analysis}\label{sec:data_analysis}

To answer our research questions (which is done in \autoref{sec:results}), we analyzed a total of \nbPapersFinalSelectionAndSnowball papers. The analysis was composed of data extraction and synthesis.

\newpage

\vspace{5pt}
\noindent\textbf{Data extraction.}
Initially, we read all the papers in full and extracted the necessary data to answer our research questions. The papers were equally divided among the authors, and periodic meetings were held for discussion. For RQ1, we extracted the independent variables of the studies, which are the formatting elements (e.g., indentation) and their levels (e.g., two-space indentation) being compared. For RQ2, for each formatting element, we collected 1) the tasks the human subjects were required to perform in the experiment, which we mapped to learning activities (see \autoref{sec:learning_activity}), 2) the dependent variables, which are the response variables (see \autoref{sec:tasks_response_variables}), 3) the results and statistical analysis, 4) the characteristics of the human subjects, and 5) the programming languages considered.

\vspace{5pt}
\noindent\textbf{Synthesis process.}
To better understand and present the results in an organized way, we performed card sorting~\citep{wood2008card} on the extracted formatting elements. With this method, we transformed the formatting elements into cards and then grouped the cards based on their similarities. This process was bottom-up, i.e., formatting elements $\rightarrow$ groups, as follows.

\vspace{5pt}
\noindent\textit{Formatting elements (factors).}
We used the formatting elements extracted from the studies as factors in the card-sorting process. Then, we created a card for each factor evaluated by the study (e.g., indentation). For synonymous factors, we created one unique card to represent them. For instance, \cite{Santos2018} evaluated whether using a beginning block delimiter in the same line of their corresponding statements (e.g., class declaration) is better than in its own line. Similarly, \cite{Arab1992} compared different presentation schemes (Peterson, Crider, and Gustafson) that define where the block delimiter should be placed. We created a unique card named ``Block delimiter location'' for these factors. The levels (e.g., in the statement line and in a separate line) of each factor compared by the studies are described and evaluated separately.

\vspace{5pt}
\noindent\textit{Groups.}
The next step was to group similar cards. In live sessions, we discussed each card and included it in a representative group. When there was no representative group for a card, a new group was created. For example, the first card we analyzed was \textit{``Vertical and horizontal spacing''}. We created the first group, named ``Group 1'', and included that card in it. The second card we analyzed was \textit{``Formatting layout''}. Such a card is not similar to the card in ``Group 1'', so we created a second group named ``Group 2'' for it. The third card we analyzed was \textit{``Indentation''}, which is similar to the card in ``Group 1'', so we included it in that group as well. After including all cards in some groups, we gave meaningful names to each group. We also split groups that seemed too generic into smaller ones. In cases where two groups were very similar and small, we combined them.

The card sorting process was initially performed by the first two authors of this paper, and the results were later refined with the collaboration of all authors in multiple live sessions. The identified groups and factors are presented in \autoref{tab:card-results} and detailed in \autoref{sec:results}.

\subsection{Differences between this study and the previous one}\label{sec:diffs}

This work is part of a bigger research project where we investigate formatting elements, coding idioms, and programming constructs that affect code legibility and readability, what they are, which ones are the best, and how researchers and developers evaluate what ``best'' means. Thus, the methodology we used in this work resembles the methodology of a previous study \citep{oliveira2020}, but the studies are different. We dedicate this section to explaining the differences between the two works.

In the previous paper, our goal was to examine and classify the tasks performed by human subjects in experiments and response variables employed by researchers to assess code legibility and readability. This paper builds upon our previous work, so much so that \autoref{sec:background} presents a summary of that paper, and aims to investigate formatting elements that were compared in the literature and which ones are more legible. Because the goals of the two studies are different, the research questions and data analysis (explained in \autoref{sec:data_analysis}) are also different.
Moreover, we changed the process of the search for papers to better match the goal of this study. We used papers found in our previous work as seed papers for defining a new search string for this work.

The remainder of the methodology, which deals with study selection (i.e., study exclusion, study inclusion, and study quality assessment), was reused or adapted from our previous work. Because some studies that were selected in our previous work are not adequate to answer the research questions of this study, we changed inclusion criteria, more specifically IC1 and IC3, to select only studies on legibility that directly compare formatting elements or styles that can be clearly identified. Finally, a subset of the \nbPapersFromAllEnginesAndSnowball documents had already been analyzed in the selection phases of our previous work, and therefore we partially reused the results from it.


\subsection{Data availability}

Raw data, such as the list of the \nbPapersFromAllEngines documents returned by our automatic search, \nbPapersFromSnowballing documents returned by snowballing, and additional details about our methodology, such as the questionnaires we used for the study quality assessment and data extraction, are available at \url{https://github.com/reydne/code-comprehension-review}.

\input{tables/card-results}

\section{Results}
\label{sec:results}

In this section, we present the results of the study. We organized the subsections by the groups of formatting elements (factors) we found in our review, which are named formatting (\autoref{sec:formatting_styles}), spacing (\autoref{sec:spacing}), block delimiters (\autoref{sec:block_delimiters}), long or complex code line (\autoref{sec:long_or_complex_code_line}), and word boundary styles (\autoref{sec:word_boundary_styles}). In each section, we address both research questions. The first research question aims to identify the formatting elements that researchers investigated in their studies about code legibility, i.e., the characteristics of the code that make it easier or harder to identify its elements. For a given group, we present the formatting elements with their corresponding levels, separated per the primary studies included in our review. \autoref{tab:card-results} presents the mapping of groups to formatting elements. With the second research question, we aim to synthesize the studies' findings related to legibility by comparing the levels of the formatting elements. For that, we consider data such as the studies' answers, statistical tests, activities performed by humans, and dependent variables. \autoref{tab:included-papers} presents an overview of the papers included in our study and summarizes the factors (from \autoref{tab:card-results}) investigated in each of them.

\input{tables/included-papers.tex}

\subsection{Formatting}\label{sec:formatting_styles}

This group gathers studies about different ways of formatting code from a global, higher-level perspective. \autoref{summary-formatting-styles} presents a summary of the results. For each factor, the table presents the references to the studies, the levels of the factors compared, the programming languages considered, the dependent variables of each study, the corresponding activities (see \autoref{sec:learning_activity}) performed by the subjects, and the main results found for each comparison of levels.

\input{tables/summary-formatting-styles}

\vspace{5pt}
\noindent\textbf{Formatting style.}
\cite{Oman1990} proposed the notion of Book Format Style for structuring source code and compared it to two well-known styles for the Pascal and C languages: the Lightspeed Pascal style \citep{johnson1988} and the Kernighan \& Ritchie style \citep{ritchie1988}. The Book Format style, as the name implies, takes inspiration from how books are structured as an approach to organizing source code. In this format, programs include a preface, table of contents, chapter divisions, pagination, code paragraphs, sentence structures, and intramodule comments, among other elements. For the comparison with the Lightspeed Pascal style, the authors asked 36 students to answer 14 multiple-choice, short-answer questions about the characteristics of a code snippet in a total of 10 minutes (both requiring the \textsf{Trace} activity), and to provide a subjective opinion about the understandability of the code snippet (\textsf{Giving an opinion}). The subjects were randomly assigned to two treatment groups, one for each formatting style, and both received the same instructions. Then, the authors assessed (i) the number of questions the students answered correctly (score of 1 to 14 points), (ii) the time they spent answering the questions (1 to 10 minutes), (iii) their performance score (number of correct answers per minute), and (iv) their subjective opinions (rating on a five-point scale). The study concluded that the Book Format style is considered a better formatting style compared to the Lightspeed Pascal style based on the significant differences in correctness score (ANOVA, $p < 0.005$), performance (ANOVA, $p < 0.01$), and subjective understandability opinion (ANOVA, $p < 0.05$), but not for time. In the comparison between the Book Format style and the Kerningham \& Ritchie style, the authors prepared a very similar experiment with 44 different students. The only difference is that the questionnaire had 10 questions instead of the 14 from the previous experiment. They assessed the same response variables and found out that the Book Format style was statistically better for correctness score (ANOVA, $p < 0.005$) and performance (ANOVA, $p < 0.005$), but there are no significant differences for time and subjective opinion.

\vspace{5pt}
\noindent\textbf{Formatting layout.}
\cite{Siegmund2017} compared a pretty-printed code layout with a disrupted layout. More specifically, they compared the code comprehension of subjects considering code snippets following common coding conventions for layout, e.g., indentation, line breaks, and correctly-placed scope delimiters, and a code snippet with defective layout, e.g., line breaks in the middle of an expression or irregular use of indentation. An interesting differentiating aspect of such a study is that it analyzed bottom-up program comprehension and the impact of beacons \citep{brooks1978} and pretty-printed layout using functional magnetic resonance imaging (fMRI). Eleven students and professionals took part in the study. First, these subjects participated in a training session in which they studied code snippets in Java, including semantic cues, to gain familiarity with them. Then, once in the fMRI scanner, the participants looked at other small code snippets to determine whether they implemented the same functionality as one of the snippets in the training session with a time limit of 30 seconds for each snippet (\textsf{Relate}). In the scanner, the level of blood oxygenation in various areas of the brain was measured. This process is called BOLD (Blood Oxygenation Level Dependent) \citep{chance1993}. That metric is a proxy to assess the activation of the area of the brain.
To evaluate the role of beacons and layout on comprehension based on semantic cues, they created four versions of the semantic-cues snippets in Java: (i) beacons and pretty-printed layout, (ii) beacons and disrupted layout, (iii) no beacons and pretty-printed layout, (iv) no beacons and disrupted layout. Finally, they extracted the Random-effects Generalized Linear Model (GLM) beta values for each participant and condition to identify differences in brain activation for each program comprehension condition. For that evaluation, we considered the study with code snippets in which there are no beacons to avoid the effect of another variable. Specifically, we avoid beacons because they are not related to formatting elements. The authors did not find significant differences in activation values in the brain areas when comparing code versions with pretty-printed layout and disrupted layout.

\subsection{Spacing}\label{sec:spacing}

This group assembles studies about different types of spacing in source code. \autoref{summary-spacing} presents a summary of the analyzed results.

\vspace{5pt}
\noindent\textbf{Indentation.}
\cite{Miara1983} conducted an experiment with 54 novice students and 32 professional developers and experienced students (i.e., people with three or more years of programming experience) to evaluate the influence of indentation levels and blocked code on program comprehension. In the experiment, the authors used the following independent variables: level of indentation (zero, two, four, and six spaces), level of experience (novice and expert), and method of block indentation (blocked and non-blocked): the latter belongs to block delimiters, so it is presented in \autoref{sec:block_delimiters}. Each subject received a program in Pascal using one of the levels of indentation and one of the methods of block indentation, accompanied by a quiz with 10 questions that required subjects to (i) select the correct answer about code characteristics (\textsf{Trace} and \textsf{Inspect}), (ii) explain what the code does (\textsf{Present}), and (iii) provide an opinion on the difficulty of the task (\textsf{Giving an opinion}). The authors considered two dependent variables: the number of questions correctly answered (score between 1-10) and the rating of the subjective opinion. The results showed that there are significant differences between novices and experts (ANOVA, $p < 0.001$) and between indentation levels (ANOVA, $p = 0.013$), where two-space indentation resulted in the best results.

\input{tables/summary-spacing}

\cite{Furman2002} also evaluated the effect of the left (no space), normal (2-4 spaces), and random indentation on code legibility. They conducted an experiment with 24 inexperienced and 18 experienced programmers in Pascal. The subjects had to understand three sorting programs, one at a time, presented in one indentation version. The authors used DISCOVERY (a tool developed by them) to measure visual metrics. This tool printed the programs to mask the code lines with the character `X' and only allowed subjects to look at one line at a time. After understanding the program, the subjects had to answer eight multiple-choice questions for each of the three sorting programs (\textsf{Trace, Inspect, and Present}). The authors evaluated the correctness of the answers, the average line-look time (i.e., time spent looking to a revealed code line), the average line-search time (i.e., time spent choosing a code line to be revealed), and the total of code lines revealed in a program. Also, the authors asked subjects to answer a subjective questionnaire (\textsf{Giving an opinion}) using the Likert scale to evaluate the difficulty of performing the task, the subjective preference, and the level of fatigue. They found a statistical difference between different indentation versions in terms of correctness. With the visual metrics, they found a statistical difference in line-look time (F-test, $p<0.05$) and in the total of revealed lines (F-test, $p<0.05$). For the line-look time, subjects spent less time in program versions with left and normal indentation than in programs with random indentation, and for the total of revealed lines, they spent less time in programs with normal indentation than in other versions. The authors did not find a statistical difference for line-search time. Finally, the subjects considered the version with normal indentation less difficult than others (F-test, $p<0.05$), as well as their subjective preference (F-test, $p<0.05$). However, they considered that version with normal indentation only was less fatigued than the random version (F-test, $p<0.05$).

\cite{Santos2018} performed a survey with 55 students and 7 professionals to investigate the impact of a set of Java coding practices on code understandability. Among other practices, subjects had to choose (\textsf{Giving an opinion}) between two alternatives for indentation level: two and four spaces. The authors evaluated the results by comparing the proportions of the votes to each alternative. The results showed that there is no statistical difference (Two-tailed test for proportion, $p=0.32$) between these alternatives.

\cite{Bauer2019} conducted an experiment with 7 developers and 15 students to investigate the influence of levels of indentation (zero, two, four, and eight spaces) on code comprehension. They designed an experiment similar to the one of~\cite{Miara1983}, with the differences that they asked subjects for the output of code snippets in an open box instead of providing subjects with multiple choices, and did not ask for a description of code functionality. They assessed code comprehension by (i) the correctness of answers (\textsf{Trace}) and (ii) the rank of perception of task difficulty (\textsf{Giving an opinion}). Differently from the work of~\cite{Miara1983}, this experiment included eye-tracking to measure visual effort using fixations, which occur when the gaze is resting on a point, and saccades, which are the transition between two fixations. Furthermore, it measured the time subjects spent providing an answer. The code snippets in this experiment were in Java.
The authors first applied Mauchly's sphericity test to evaluate which test was more adequate for each dependent variable, i.e., one-way ANOVA with repeated measures or Friedman's test. The results of this experiment were: the correctness of answers (Friedman's test, $p=0.36$), the log-transformed response time (ANOVA, $p=0.72$), perceived difficulty (Friedman's test, $p=0.15$), fixation duration (ANOVA, $p=0.045$), fixation rate (Fridman test, $p=0.06$), saccadic amplitude (ANOVA, $p=0.18$). Despite the $p<0.05$ for fixation duration, the authors could not confirm this difference with a post hoc test. Therefore, unlike in the previous study, there was no statistically significant difference between indentation levels.

\vspace{5pt}
\noindent\textbf{Appropriate use of indentation with blocks.}
In the work by \cite{langhout2021}, a partial replication of the study presented by~\cite{Gopstein2017} (further explained in \autoref{sec:block_delimiters}) was conducted, focusing on the Java language instead of C/C++. These papers measure the impact of atoms of confusion in code understandability. The authors recruited 132 novice developers for an experiment conducted in two parts, in which participants had to (i) predict the output of code snippets with and without atoms (\textsf{Trace}) and (ii) give their opinions about which ones are confusing among two functionally-equivalent code snippets, one with an atom of confusion and another without (\textsf{Giving an opinion}). For the first part, the authors calculated the odds ratio of subjects correctly predicting the output of the code snippets as a measure of effect size. 
Two atoms of confusion related to inappropriate use of indentation were investigated: the Remove Indentation atom and the Indentation atom. In both cases, there exist indented statements after the end of a block, which makes it look like the statements are part of the block. In the latter case, the incorrectly indented statement is preceded by curly braces, whereas in the former, it is not.
The authors found out that Remove Indentation is associated to misunderstanding (odds ratio $= 56.21$, $p<0.05$), and for 57.1\% of the subjects, this atom is unfavorable for legibility. There was no statistical difference in answers for Indentation, although 72.7\% of the subjects considered it confusing.

\vspace{5pt}
\noindent\textbf{Vertical and horizontal spacing.}
\cite{Love1977} evaluated the use of paragraphed vs. unparagraphed code, i.e., the disciplined use of vertical and horizontal spacing to organize the code in text-like paragraphs. In the paragraphed version, the source code contains blank lines between different instructions and indentation to represent blocks. Unlike, there are no blank lines or indentation in the unparagraphed version. The author experimented with 19 undergraduate and 12 graduate students to investigate the effect of program indentation (paragraphed or unparagraphed) and control flow (simple or complex) on code comprehension. The subjects were asked to memorize a program in Pascal in 3 minutes, rewrite the program in 4 minutes (\textsf{Memorize}), and describe the program functionality (\textsf{Present}). The reconstructed programs were scored based on the percentage of lines of source code correctly recalled in the proper order. The descriptions of the program functionality were rated with a 5-point scale, where one point means that a description has nothing right about the program functionality and five points means that a description shows a complete understanding of it. The author did not find a statistically significant difference between paragraphed and unparagraphed programs, considering the correctness scores in the \textsf{Memorize} activity (ANOVA, $p=0.79$) and in the \textsf{Present} activity (ANOVA, $p=0.6$) for both undergraduate and graduate students.

\vspace{5pt}
\noindent\textbf{Vertical spacing between related instructions.}
Another coding practice analyzed by~\cite{Santos2018} pertains to the use of vertical spaces. More specifically, the authors asked the subjects (by presenting code examples) whether blank lines must be used to create a vertical separation between related instructions (\textsf{Giving an opinion}). The results did not show a statistically significant difference (Two-tailed test for proportion, $p=1.0$) between subjects who agree and disagree with this practice.


\vspace{5pt}
\noindent\textbf{Blank space around operators and parameters.}
\cite{sampaio2016} investigated the importance of teaching a set of best practices for code understandability. For this, the authors conducted a survey to ask object-oriented programming teachers to provide a subjective rating on a 5-point Likert scale about the importance of coding practices for code understandability (\textsf{Giving an opinion}). The authors considered that practices with a median rating greater than 3 are relevant to code understandability. Four practices related to legibility were investigated, among others.
The practices related to spacing, which are the ones relevant to this section, are about blank space around operators and arguments/parameters: \textit{``The assignment operator include blank space before and after''} and \textit{``Blank space between the arguments/parameters of functions''}. The authors found out that the first ($median = 3$, $range = 4$, $interquartile = 2$) and the second ($median = 2$, $range = 4$, $interquartile = 2$) practices are not relevant for legibility.

\subsection{Block Delimiters}\label{sec:block_delimiters}

This group gathers studies about different types of block/scope delimiters. \autoref{summary-block-delimiters} presents a summary of the analyzed results.

\input{tables/summary-block-delimiters}

\vspace{5pt}
\noindent\textbf{Block delimiter location.}
\cite{Miara1983} compared two styles of block indentation: (i) blocked, where the code within a block is at the same indentation level as the block delimiters (begin-end), and (ii) non-blocked, where the code within the block appears at least one more indentation level to the right of the block delimiters.
This comparison is different from what has been previously reported about indentation (\autoref{sec:spacing}) because this one evaluated when the indentation of the block starts (i.e., in the line of the block delimiter itself or in the first statement after the block delimiter), while the other one focuses solely on the indentation level. The design of this study was explained in~\autoref{sec:spacing}. The analysis of variance (ANOVA) of the quiz scores and the program ratings showed no significant effect with the blocked and non-blocked styles.

\cite{Arab1992} compared three combinations of the use of block delimiters in the same line as their associated statements with the use of block delimiters in a separate line. More specifically, they evaluated three different code presentation schemes in Pascal that propose different patterns to place the block delimiters: (i) Peterson's scheme: BEGIN and END in their own lines~\citep{peterson1977}; (ii) Crider's scheme: BEGIN and END in the same line of the last statement or declaration~\citep{crider1978}; and (iii) Gustafson's scheme: BEGIN in the same line as the last statement but END in its own line~\citep{gustafson1979}. The study consisted of an opinion survey with 30 subjects (22 novices and 8 experts), who were asked to classify the three schemes in descending order (\textsf{Giving an opinion}). The results showed that the block delimiters in their own lines, i.e., Peterson's scheme, was chosen as the best by most participants. However, they did not apply any statistical test to evaluate the results. In a similar vein, \cite{Santos2018} asked subjects whether they prefer to have Java's opening curly braces in their own lines or in the same line as their corresponding statements (e.g., class declaration). They found a statistically significant difference (Two-tailed test for proportion, $p=0.006$) between subjects who preferred the first alternative and subjects who preferred the second alternative, where most participants consider that opening braces in their own lines are easier to read, similar to results of \cite{Arab1992}.

\vspace{5pt}
\noindent\textbf{Block delimiter style.}
In the study reported by \cite{Sykes1983}, different styles of scope delimiters in Pascal were investigated: (i) ENDIF, which uses \texttt{IF-ENDIF} and \texttt{WHILE-ENDWHILE} scope delimiters for the bodies of conditional and iterative statements, respectively; (ii) REQ-BE, which always uses \texttt{BEGIN-END} scope delimiters for simple and compound statements within a conditional structure; and (iii) BE, which uses \texttt{BEGIN-END} scope delimiters only for compound statements. The study employed a questionnaire where most of the questions asked subjects to determine the values of variables by simulating the execution of program segments in Pascal with initial values given in the questions (\textsf{Trace}). In total, 36 students participated in that study, and they had 25 minutes to finish the questionnaire. The subjects were divided into advanced (at least two years of programming experience) and intermediate. The authors evaluated the responses to the questionnaire considering the type of delimiter and the experience of the subjects. ANOVA indicated that experience was significant ($p=0.005$), and advanced subjects did much better than intermediate ones. Also, a paired t-test (considering an $\alpha$ = 0.1) showed that the subjects performed better using the ENDIF delimiters than using the REQ-BE  ($p=0.074$) and BE ($p=0.073$) delimiter styles.

\vspace{5pt}
\noindent\textbf{Block delimiter visibility.}
Also, in the work of \cite{Sykes1983}, we found an evaluation of block delimiter visibility in Pascal, i.e., whether omitted or present block delimiters impact code legibility. We limited the comparison analysis between REQ-BE and BE to evaluate the block delimiter visibility. The details of the study are explained in the previous paragraph. The authors did not find a difference between the REQ-BE and BE delimiter styles (t-test, $p=0.955$).

The work of \cite{sampaio2016} (introduced in \autoref{sec:spacing}) evaluated the relevance of the practice ``\textit{use \{ and \} to enclose the statements in a loop}'' for code legibility, in the context of Java code. They found that this practice is not relevant ($median = 2$, $range = 4$, $interquartile = 2$) for legibility.

\cite{Gopstein2017} compared the use of omitted and present block delimiters in the specific context of small C programs. The authors conducted a broad-scoped experiment\footnote{Arguably, it was an Internet-based survey, but with enough controls in place that we feel it is more appropriate to call it an experiment.} with 73 programmers with three or more months of experience in C/C++ to evaluate small and isolated patterns in C code, named atoms of confusion, that may lead programmers to misunderstand code. They asked subjects to analyze tiny programs ($\sim$8 lines), where the control version contained a single atom of confusion candidate, and the treatment version contained a functionally-equivalent version where the atom does not exist. For each program, the participants had to predict its output (\textsf{Trace}). The authors compared the correctness of the answers of the version with the atom of confusion and their counterparts without the atoms. One of these atoms is called \textit{Omitted Curly Braces}. In C programs, this is used when the body of an \texttt{if}, \texttt{while}, or \texttt{for} statement consists of a single statement. The study found a statistically significant difference (McNemar’s test adjusted using Durkalski correction, $p<0.05$) between the programs where the curly braces were omitted and the ones where they were not. Subjects analyzing programs with omitted curly braces made more mistakes. \cite{langhout2021} (introduced in \autoref{sec:spacing}) also evaluated the atom \textit{Omitted Curly Braces}, but in Java code. They found a statistically significant difference (odds ratio $= 4.62$, $p<0.05$), indicating that omitting curly braces is associated with misunderstanding. Furthermore, 93.3\% of the subjects agreed that a code snippet with this atom is hard to read.

Similarly, \cite{Medeiros2019} investigated what they call misunderstanding code patterns, which are very similar to atoms of confusion. One study was a survey with 97 developers, where the subjects should determine the negative or positive impact of using one code snippet version containing one misunderstanding pattern instead of a functionally-equivalent alternative without that pattern on a 5-point Likert scale (\textsf{Giving an opinion}). In another study, they submitted 35 pull requests suggesting developers remove several misunderstanding patterns. This study focused on the C language.
In this section, we analyze the results of the misunderstanding pattern named \textit{Dangling Else}, which happens when an \texttt{if} statement without braces contains one \texttt{if}/\texttt{else} statement without braces. The reader of the code needs to know that the \texttt{else} clause belongs to the innermost \texttt{if} statement. The first study found that \textit{Dangling Else} is perceived as negative by 71.73\% of the developers. In the second study, they submitted 10 pull requests to remove instances of the \textit{Dangling Else} pattern, where two of them were accepted, one was rejected, three were not answered, and four were ignored by developers (according to the authors, due to the pattern instances being in third party or deprecated code).

\subsection{Long or Complex Code Line}\label{sec:long_or_complex_code_line}

This group comprises studies that evaluate elements related to the size and complexity of code lines. \autoref{summary-long-complex-line} presents a summary of the analyzed results.

\vspace{5pt}
\noindent\textbf{Line length.}
In the study of \cite{Santos2018} (details about the methodology are in \autoref{sec:spacing}), the subjects were asked if they agree that the practice ``\textit{line lengths not exceeding 80 chars}'' has a positive impact on code legibility. The authors found out that students and professionals find code snippets containing lines of up to 80 characters more legible than code snippets where there are longer lines (Two-tailed test for proportion, $p<0.001$). 

\vspace{5pt}
\noindent\textbf{Statements per line.}
Three studies tackle the question of whether lines of code including a single statement are more legible than ones with multiple statements. \cite{sampaio2016} evaluated the relevance of the practice of \textit{``breaking the line after semicolon''} for code legibility (the methodology is detailed in \autoref{sec:spacing}). The authors found that this practice is relevant for code legibility ($median = 4$, $range = 3$, $interquartile = 1.5$).

\cite{Santos2018} found out that students and professionals agreed with the practice ``\textit{avoid multiple statements on a same line}'' (Two-tailed test for proportion, $p<0.001$), i.e., the subjects prefer code snippets where lines of code do not include multiple statements (details about the methodology are in \autoref{sec:spacing}). As discussed before, the paper of \cite{Medeiros2019} (\autoref{sec:block_delimiters}) introduced and evaluated what the authors call misunderstanding code patterns. Among the studied patterns, one specifically refers to a scenario where multiple variables are initialized on the same line. The study asked subjects their opinion about the use of this pattern (\textsf{Giving an opinion}). The authors found out that the use of \textit{multiple initializations on the same line} was neither negative nor positive for most of the subjects. In the second study, they submitted one pull request to remove this pattern, which was rejected by developers. 

\input{tables/summary-long-complex-line}

\subsection{Word Boundary Styles}\label{sec:word_boundary_styles}

The fifth and last group we have identified comprises studies of word boundary styles for identifiers.
\autoref{summary-word-boundary-styles} presents a summary of the analyzed results.

\vspace{5pt}
\noindent\textbf{Identifier style.}
\cite{Sharif2010} investigated the effect of the camel case and underscore identifier styles on code legibility. They compared the identifier styles using phrases formed by two and three words from code, i.e., phrases that are likely to be in source code (e.g., ``start time''), and non-code, i.e., phrases that are not likely to be in source code (e.g., ``river bank''). In the experiment, 15 students had to read a phrase and, on the next screen, choose an identifier (from four choices) that exactly matches the phrase they just saw (\textsf{Memorize}), and answer it verbally. Only one of the choices is correct, and the rest are distracters that change the beginning, middle, or end of the identifier. It was repeated for eight phrases. The authors measured the correctness, time, and visual effort for each phrase. They found no difference between identifier styles for correctness.
Using a simple linear mixed model, they found a significant difference ($p=0.0001$) in time where subjects took 13.5\% longer for camel-cased identifiers than underscored. For visual effort, they employed two metrics: fixation rate and average fixation duration. They were measured when the subjects looked at the correct answer and the distracters. Using Wilcoxon test, they only found a significant difference in the average fixation duration (correct: $p=0.015$; distracters: $p=0.026$), where the distribution showed that camel-cased identifiers required a higher average duration of fixations than underscored.

\cite{Binkley2013} performed five studies to evaluate the effect of identifier style on code comprehension. Only three are related to source code: Where's Waldo study, Eye Tracker Code study, and Read Aloud study. In the first study, 135 programmers and non-programmers attempted to find all occurrences of a particular identifier in code fragments (\textsf{Inspect}) written in C and Java. The authors measured the number of lines where the subjects misread occurrences of the identifier and the time spent on that task for the subjects that found all identifier occurrences. Using a Linear mixed-effects regression, they found a significant difference ($p=0.0293$) in favor of camel-cased identifiers, to an estimated 0.24 fewer missed lines than with underscored ones. For the time, it was marginally significant ($p=0.0692$) and indicated that camel-cased identifiers took, on average, 1.2 fewer seconds to be found than underscored ones.

\input{tables/summary-word-boundary-styles}

In the second study, they asked 15 programmers (undergraduate and graduate students) to study two C++ code snippets, reproduce them, and answer questions by selecting the identifiers they remember from the code (\textsf{Memorize}). The authors considered the correctness of the answers. They also used an eye tracker to measure eye fixations and gaze duration, which is the duration of a fixation within a specific area of interest (boxes around the identifiers with some extra padding). The authors compared differences in visual effort using two pairs of similar identifiers, \textit{row\_sum} and \textit{colSum}, and \textit{c\_sum} and \textit{rSum}. They found out that \textit{row\_sum} required significantly more fixations than \textit{colSum} (1-tailed, $p=0.007$), and that its average gaze duration was 704ms longer than that of \textit{colSum} (1-tailed, $p=0.005$). No significant differences were found between \textit{c\_sum} and \textit{rSum}. Furthermore, they did not find a statistical difference (Linear mixed-effects regression, $p=0.451$) for correctness.

Finally, the third study asked 19 programmers to summarize (i.e., explain each step of the code) a Java code snippet (\textsf{Present}) verbally. The subjects were a subset of the Where's waldo study group in at least their second year of university. The authors collected the amount of time that each subject spent reviewing the code before summarizing it and the correctness of the answer, which was assessed on a 10-point Likert scale by the authors. They did not find a statistical difference between camel case and underscore for time (Linear mixed-effects regression, $p=0.6129$) or correctness (Linear mixed-effects regression, $p=0.3048$). 

\subsection{Addressing the two research questions} 

The two research questions that this work aims to answer are:  

\begin{description}
\item[RQ1] \rqone
\item[RQ2] \rqtwo
\end{description}

Based on the results of our study, for RQ1, we have identified \nbPapersFinalSelectionAndSnowball scientific papers in which researchers compared alternative levels of formatting elements with human subjects. We found \nbFactors factors (i.e., formatting elements), which are about code formatting (e.g., formatting layout), code spacing (e.g., indentation), block delimiters (e.g., block delimiter location), long or complex code lines (e.g., line length), and word boundary styles (i.e., identifier style). For the \nbFactors factors, researchers examined \nbLevels different levels (e.g., 2 and 4 levels of indentation) in \nbComparisons comparisons (e.g., one comparison considered 0, 2, 4, and 6 indentation levels).

For RQ2, researchers found statistically significant results in \numberToBeChecked{17} comparisons and no significant results in \numberToBeChecked{10} comparisons.
Considering the significant differences, it was found that the best alternative levels for \numberToBeChecked{four} factors were:
book format style (cf. Lightspeed Pascal and Kernighan \& Ritchie styles), 
appropriate use of indentation with blocks (cf. inappropriate use), 
the use of \texttt{ENDIF/ENDWHILE} for block delimiters (cf. the use \texttt{BEGIN-END} in all blocks and the use \texttt{BEGIN-END} for compound statement blocks), and 
the practice that line lengths should be kept within the 80-character limit (cf. line lengths that exceed such a limit).
In other comparisons where significant differences were found, there were divergent results.
In one case, more than one comparison of \numberToBeChecked{the same factor} found statistically significant results but in favor of different levels. This is the case for the factor identifier style, for which one study found that underscore is the best alternative for identifier style while another study found that camel case is better than underscore.
In other cases of divergent results, which involved \numberToBeChecked{four} factors, some comparisons found significant differences in favor of some level, but other comparisons found no significant differences. This is the case for the factors indentation, block delimiter location, block delimiter visibility, and statements per line.
Finally, for \numberToBeChecked{four} factors (i.e., formatting layout, vertical and horizontal spacing, vertical spacing between related instructions, and blank space around operators and parameters), no statistically significant results were found for their levels.

\section{Discussion}

In this section, we discuss aspects of this study that pertain to more than one of the investigated papers. We also highlight gaps we have identified in the literature and potential directions for future work.

\subsection{Contrasting empirical results with existing coding style guides}

In this section, we discuss the results of this study in the light of five existing style guides for Java (Google Java Style Guide\footnote{\url{https://google.github.io/styleguide/javaguide.html}, last access May 22, 2023.}), JavaScript (AirBNB JavaScript Style Guide\footnote{\url{https://github.com/airbnb/javascript}, last access May 22, 2023.}), Python (Style Guide for Python Code\footnote{\url{https://peps.python.org/pep-0008/}, last access May 22, 2023.}), C (Linux Kernel Coding Style\footnote{\url{https://www.kernel.org/doc/html/v4.10/process/coding-style.html}, last access May 22, 2023.}), and Pascal (Free Pascal\footnote{\url{https://wiki.freepascal.org/Coding_style}, last access May 22, 2023.}), and contrast their recommendations against the findings of this study.

\vspace*{0.2cm}
\noindent
\textbf{Formatting.} We found three comparisons of formatting styles, but only two of them showed statistically significant results. In particular, the Book Format style~\citep{Oman1990} exhibited better results compared to the Lightspeed Pascal and Kernighan \& Ritchie styles. However, one of these results is from a study that used Pascal, which is not a popular programming language in 2022\footnote{\url{https://www.tiobe.com/tiobe-index/}, last access May 22, 2023.}\footnote{\url{https://redmonk.com/sogrady/2022/03/28/language-rankings-1-22/}, last access May 22, 2023.}, and at the time this comparison was performed, formatting tools were still scarce. Such results may not be applicable in our current context because considerable advances have been made in programming languages and tools. Furthermore, no existing coding style guide supports the use of the Book Format style.

\vspace*{0.2cm}
\noindent
\textbf{Spacing.} Most studies that evaluated spacing did not produce statistically significant results. The results of one study~\citep{Love1977} indicate that the disciplined use of vertical and horizontal spacing to organize code into text-like paragraphs is not relevant. We hypothesize that this may stem from a lack of statistical power: different spacing approaches are not likely to yield big effect sizes, and the power required to detect such small effect sizes implies large sample sizes (\autoref{sec:power}). 
Two studies~\citep{Santos2018,Bauer2019} did not find a significant difference between different indentation levels, e.g., two versus four spaces, in Java code. An older study~\citep{Miara1983} comparing zero, two, four, and six spaces in Pascal code snippets found that two spaces exhibited the best result. The coding style provided by Free Pascal recommends two spaces for indentation. Although \cite{Miara1983}'s finding is for Pascal, it is also consistent with some coding guides such as the Google Java Style Guide, the AirBNB JavaScript Style Guide, and the Style Guide for Python Code. Others, such as the Linux Kernel Coding Style, recommend eight spaces and explicitly argue against two or four spaces.
Currently, the developer community uses spacing patterns, both vertical and horizontal, because they believe in their usefulness.

\vspace*{0.2cm}
\noindent
\textbf{Block delimiters.} The majority of the comparisons related to block delimiters (\numberToBeChecked{7/10 comparisons}) showed statistically significant results. They evaluated the location, visibility, and style of block delimiters. The studies (in Pascal, C, C++, and Java) suggest that block delimiters are relevant to legibility, and should be visible and stay in their own line. The Google Java Style Guide is consistent with this result, suggesting the use of braces even when the block is empty or contains only a single statement (block delimiter visibility), but not about the placement of braces (block delimiter location). The Linux Kernel Coding Style agrees with this for functions but disagrees for statements. In addition, the AirBNB JavaScript Style Guide prescribes that the opening brace of a statement or declaration should be placed in the same line as the statement or declaration and not in a separate line.

Some languages allow omitting the block delimiters in some cases, e.g., when the block consists of only one line, to make the code less verbose. The Linux Kernel Coding Style argues in favor of this practice. The Google Java Style Guide and the AirBNB JavaScript Style Guide argue in the opposite direction. The results obtained by \cite{Gopstein2017} suggest that omitting braces may be bug-prone. Furthermore, using names to block delimiters that indicate the context of the block seems to increase code comprehension. It becomes apparent when the code has nested blocks, where the information about the block in the name of delimiters could avoid misunderstanding. To the best of our knowledge, no programming language in widespread use supports this approach.

\vspace*{0.2cm}
\noindent
\textbf{Long or complex code line.} One study investigated and found that keeping line lengths within 80 characters in Java is considered positive for legibility \citep{Santos2018}. On the other hand, \cite{sampaio2016}, \cite{Santos2018}, and \cite{Medeiros2019} investigated the use of one vs. multiple statements per line and obtained different results. \cite{sampaio2016} and \cite{Santos2018} found out that one statement per line in Java is preferred by the participants, but \cite{Medeiros2019} did not find a significant difference between one or more statements per line in C. In these three studies, only the subjects' opinions were used as the dependent variable. The Style Guide for Python Code and Linux Kernel Coding Style agree with the 80-character limit. The Google Java Style Guide recommends one statement per line but recommends 100-character lines. The AirBNB JavaScript Style Guide also establishes 100 characters as the limit for lines, but it makes no prescription about the number of statements per line.

\vspace*{0.2cm}
\noindent
\textbf{Word boundary styles.} The two studies that investigated word boundary styles for identifier names revealed divergent results. While \cite{Binkley2013} found that camel case is a positive standard for legibility in comparison to snake case (a.k.a. underscore) in Java and C, \cite{Sharif2010} found that snake case is the best alternative (no specific programming language was considered). The camel case style is adopted in the examined coding style guides for Java and JavaScript. Python uses camel case for class names and for method names \textit{``where that’s already the prevailing style [...] to retain backwards compatibility''}. For most names, it suggests the use of snake case, with additional trailing or leading underscores for special identifiers. The Linux Kernel Coding Style explicitly argues against camel case and prescribes the use of snake case.

\subsection{Statistical power of the analyzed studies}\label{sec:power}

The power of a statistical test describes the probability that a statistical test will correctly identify a genuine effect~\citep{Ellis2010}. In other words, a statistical test with sufficient power is unlikely to accept a null hypothesis when it should be rejected. A scenario where a null hypothesis is rejected when it should not have been rejected is called a Type II error~\citep{cohen1992}. The calculation of the sample size required to achieve a certain level of statistical power is called power analysis. It depends on the significance criterion, the sample size, and the population effect size. 

For \numberToBeChecked{nine} of the comparisons we have investigated, it was not possible to reject the null hypothesis. However, it is not clear whether these results stem from the absence of a statistically significant difference or from a lack of power. 
Among the analyzed studies, only the works of \cite{Gopstein2017} and \cite{Bauer2019} report a concern with this aspect. In both cases, the authors have calculated the sample sizes that would yield what they considered an acceptable level of statistical power. Null results reported by other studies are difficult to judge because they have no associated confidence level. This is due to the lack of power analysis and reporting of effect sizes, and the generally low sample sizes. These low sample sizes imply that these studies are only able to detect large effect sizes. The study of \cite{Siegmund2017} presents an illustrative example. It attempted to compare the differences in brain activation of subjects exposed to pretty-printed code and code whose layout is disrupted. Since this study involved only 11 participants, its power is very low, and it can only detect statistically significant differences if the effect size is large. 

Another example is the pair of studies performed by \cite{Miara1983} and \cite{Bauer2019}. Both compared different indentation levels, with the latter being inspired by the former. On the one hand, the study of Miara et al. involved 86 subjects and found a statistically significant difference between different indentation levels and also between novices and experts. On the other hand, the study conducted by Bauer et al. had only 22 participants and could not find statistically significant differences. Although still not common in Software Engineering, meta-analyses~\citep{Ellis2010} could be leveraged to combine the results of these different studies. In this manner, it would be possible to identify subtler differences if they exist. 

\subsection{Limitations of our study and of existing studies} 

The studies analyzed in this work were conducted between \numberToBeChecked{1977 and 2021} and covered several formatting elements. Nevertheless, there are other aspects beyond what we captured. For example, in our work, we did not consider studies that investigate the influence of typography, colors, contrast, or dynamic presentation of program elements on the ability of developers to identify program elements. In summary, aspects that fall outside what can be tinkered with at the source code level, in an ASCII text editor, are beyond the scope of this work. Investigating aspects that go beyond these limits is left for future work.

Even within the strictly-defined boundaries we adhere to, much is still unknown. This study highlights the limitations in our current understanding of the impact of formatting elements on code legibility. We have found only \nbPapersFinalSelectionAndSnowball papers containing direct comparisons between alternative levels of formatting elements and styles out of \nbPapersFromAllEnginesAndSnowball examined documents. Replications, even partial ones, are almost non-existent. This suggests that this area of research is \textbf{immature} and can greatly benefit from new studies.

As pointed out when discussing statistical power, many previous studies are \textbf{inconclusive}, and it is still not clear if well-established practices actually have any effect on legibility. For example, we identified only one paper that reports on the comparison of different types of block delimiters~\citep{Sykes1983}, and that paper was published 40 years ago, at a time when software development was very different from what it is today. 
Another example is indentation. Although many style guides provide directives about how to indent code, most of the studies on the topic found inconclusive results (\autoref{sec:spacing}). Notwithstanding, the importance of the topic for software developers in practice emphasizes that more investigation is required.
One point that can be raised against this kind of study is whether such choices matter at all, especially for experienced developers, since these can be seen as minute details. Previous work~\citep{Stefik2013,Gopstein2017,Gopstein2018,Medeiros2019} suggests that they do, though, even in very mature, high-complexity projects~\citep{Gopstein2018}.

Some of the studies we analyzed are \textbf{outdated}. These \nbOldPapers papers were published 30 or more years ago. Software development was very different back then. For example, object-oriented programming was a relative novelty, the world wide web had been public for just one year, and mobile devices did not exist for the general public. In addition, the main programming language examined by these old studies, Pascal, is rarely used in practice nowadays. These studies investigated aspects that are relevant to contemporary software development, such as indentation and the use of block delimiters. However, they do that in a scenario that is disconnected from current practice and software development methods and tools.

The learning activities involved in the analyzed studies (\autoref{sec:learning_activity}) required participants to \numberToBeChecked{\textsf{Memorize}, \textsf{Inspect}, \textsf{Trace}, \textsf{Present}, and \textsf{Relate}, plus the task of \textsf{Giving an opinion}}.
The learning activities were modeled in two semi-independent dimensions by~\cite{fuller2007developing} and extended by~\cite{oliveira2020}. Each dimension defines hierarchical linear levels where a deeper level requires the competencies from the previous ones. This means that some activities require more competencies than others. The majority of the comparisons of formatting element levels we found in the literature were based on the \textsf{Giving an opinion} task, which is not modeled because it is not a learning activity. \textsf{Trace} is the most used learning activity, but it does not even require an intermediate level of competencies. The activity used in the analyzed studies that requires more competencies is \textsf{Relate}, which was used for only one formatting element evaluation. More generally, the analyzed studies focused on ``lower'' cognitive skills according to the model of \cite{fuller2007developing}, emphasizing their \textbf{narrow scope}. Program comprehension aims to support other activities such as performing maintenance, fixing bugs, and writing tests. The absence of studies requiring participants to conduct these activities points to a possible direction for future work.

Since our ultimate goal is to create a coding style guide based on empirical evidence, an analysis of the programming languages used in the experiments of the primary studies included in our review is necessary. \numberToBeChecked{Five} studies used the Java language~\citep{sampaio2016,Siegmund2017,Santos2018,Bauer2019,langhout2021}. Another \numberToBeChecked{five} studies used the Pascal language~\citep{Love1977,Miara1983,Sykes1983,Arab1992,Furman2002}. \numberToBeChecked{Two} more targeted the C or C++ languages~\citep{Gopstein2017,Medeiros2019}. The paper by~\cite{Oman1990} used the languages C and Pascal, and the paper by~\cite{Binkley2013} used the languages C and Java. \cite{Sharif2010} did not specify a programming language. This characteristic of the studies makes it impossible to generalize the results of our study and emphasizes that very few languages have been targeted by previous work. For example, none of these studies examined a scripting language, such as Python or JavaScript, a functional programming language, e.g., Haskell or Elixir, or even more contemporary multiparadigm languages that receive strong influence from functional languages, e.g., Kotlin or Swift.

\section{Threats to validity}

In this section, we discuss the threats to the validity of this study. 
\\[0.2cm]
\noindent
\textbf{Construct validity.} Our study was built on the selected primary studies, which stem from the search and selection processes. We only used three search engines for our automatic search. Other engines, such as Springer and Google Scholar, could return different results. However, the majority of our seed studies were indexed by ACM and IEEE, and then we used Scopus to expand our search. Moreover, while searching on the ACM digital library, we used the \textit{Guide to the Computing Literature}, which retrieves resources from other publishers, such as Springer. Finally, we avoided Google Scholar because it returned more than \numberToBeChecked{17 thousand} documents for our search string, and we did not have the resources to analyze so many papers.
\\[0.2cm]
\noindent
\textbf{Internal validity.} This study was conducted by multiple researchers. We understand that this could pose a threat to its internal validity since each researcher has a certain knowledge and way of conducting her research activities. However, each researcher conducted her activities according to the established protocol, and periodic discussions were conducted between all researchers.
Moreover, a few primary studies do not report in detail the tasks the subjects perform. Because of that, we might have misclassified these studies when mapping studies to task types.
Finally, comparing different studies poses some threats due to the many differences in their setups, such as materials, programming languages, number of subjects, the subjects' expertise, and tasks to be performed by the subjects. It is not feasible to consider all differences in the studies' setups to compare the studies; some of them do not even provide fine-grained information, such as the size of source code snippets used in the experiments. In our study, however, we mitigated the threat of directly comparing studies by considering the activities performed by subjects in the experiments (which require different levels of cognitive skills), the dependent variables of the studies, the programming languages, the number of subjects, and statistical power.
\\[0.2cm]
\noindent
\textbf{External validity.} Our study focuses on studies that report alternative ways of formatting code, considering low-level aspects of the code. Our findings might not apply to other kinds of works that evaluate code legibility or readability.

\section{Conclusion}

In this paper, through a systematic review of the literature, we investigated what formatting elements have been studied and which levels of formatting elements were found to have a positive impact on code legibility when compared to functionally equivalent ones in human-centric studies. We present a comprehensive categorization of the elements and levels found. In addition, we analyzed the results considering the subjects, activities, dependent variables, and programming languages involved in the studies.

We identified \nbFactors factors of formatting elements, which were categorized into \numberToBeChecked{five} groups: formatting, spacing, block delimiters, long or complex code lines, and word boundary styles. For \numberToBeChecked{four} factors, the results showed that the levels book format style, appropriate use of indentation with blocks, the use of \texttt{ENDIF/ENDWHILE} for block delimiters, and the practice that line lengths should be kept within the limit of 80 characters exhibited positive impact on code legibility. Also, for other \numberToBeChecked{four} factors, some studies found significant results while others did not find. Furthermore, there were contradictory results in the best alternative level for \numberToBeChecked{one} factor. Finally, no statistically significant results were found for other \numberToBeChecked{four} factors.

Although we found some positive results, there are only a few suggestions that can be made to developers based on our findings. The limitations of the literature (i.e., few studies, narrow-in-scope studies, outdated studies, and inconclusive results) and, consequently, of our work, limit us toward our bigger goals: to provide developers with guidelines so that they can choose the best ways to format their code, and to help other researchers in developing automated support, e.g., linters and recommendation systems, to aid developers during programming activities and make their code more legible. Our systematic literature review allowed us to identify and discuss those limitations in this paper. Finally, these limitations call for more research and replication studies because there is much to be understood about how formatting elements influence code legibility before the creation of guidelines and automated aids to help developers.

Our study focused only on code formatting elements. Nonetheless, other studies on code comprehension investigated aspects related to the structural and semantic characteristics of the code. Thus, we plan to investigate what these characteristics are and which of them are positive for the understanding of the code in future work.

\section*{Acknowledgments}

We thank the anonymous reviewers for their valuable feedback that made us improve this paper. This work was partially supported by
the Federal Institute of Education, Science, and Technology of Pernambuco,
the Foundation for Support of Science and Technology of Pernambuco State,
the Swedish Foundation for Strategic Research under the TrustFull project,
and INES 2.0 (FACEPE PRONEX APQ 0388-1.03/14 and APQ-0399-1.03/17, and CNPq 465614/2014-0).

{\small
\bibliographystyle{elsarticle-num-names} 
\bibliography{references}
}
\end{document}

%% file: setup.tex
\usepackage{hyperref}
\hypersetup{pdfauthor={Name}} 

\usepackage{tabularx}
\usepackage{booktabs}
\usepackage{tabu}
\usepackage{multirow}
\usepackage{makecell}

\usepackage{xcolor}
\usepackage{xspace}
\newcommand*{\tabindent}{ \hspace{3mm}}
\usepackage{pifont}
\newcommand{\cmark}{\ding{51}}%
\definecolor{verylightgray}{rgb}{0.93, 0.93, 0.93}

%% file: commands.tex
\newcommand{\nbSeedPapers}{\numberToBeChecked{11}\xspace}
\newcommand{\nbPapersACM}{\numberToBeChecked{1,520}\xspace}
\newcommand{\nbPapersIEEE}{\numberToBeChecked{979}\xspace}
\newcommand{\nbPapersScopus}{\numberToBeChecked{3,458}\xspace}
\newcommand{\nbPapersFromAllEngines}{\numberToBeChecked{3,999}\xspace}
\newcommand{\nbPapersAfterExclusionCriteria}{\numberToBeChecked{550}\xspace}
\newcommand{\nbPapersAfterInclusionCriteria}{\numberToBeChecked{14}\xspace}
\newcommand{\nbPapersFinalSelection}{\numberToBeChecked{14}\xspace}
\newcommand{\nbOldPapers}{\numberToBeChecked{5}\xspace}
\newcommand{\nbFactors}{\reviewed{13}\xspace}
\newcommand{\nbLevels}{\numberToBeChecked{33}\xspace}
\newcommand{\nbComparisons}{\numberToBeChecked{27}\xspace}

\newcommand{\nbPapersFromSnowballing}{\reviewed{915}\xspace}
\newcommand{\nbPapersSnowballingAccepted}{\reviewed{one}\xspace}
\newcommand{\nbSnowballingLevels}{\reviewed{one}\xspace}

\newcommand{\nbPapersFinalSelectionAndSnowball}{\reviewed{15}\xspace}
\newcommand{\nbPapersFromAllEnginesAndSnowball}{\reviewed{4,914}\xspace}

\newcommand\reviewed[1]{\textcolor{black}{#1}}
\newcommand\numberToBeChecked[1]{\textcolor{black}{#1}}

\usepackage{todonotes}

%% file: tables/learning-activities.tex
\begin{table*}[t!]
    \caption{Learning activities from our previous study \citep{oliveira2020}, extended from \cite{fuller2007developing}---Inspect and Memorize are not in the original taxonomy. Opinion is not included because it is not directly related to learning. We took the definitions directly from \cite{fuller2007developing}, except when a definition is not applicable to our code understanding context. The examples are either extracted from tasks from the primary studies of our previous work \citep{oliveira2020} or general ones when no task involved that activity.}
    \label{tab:learning-activities}
    \centering
    \scriptsize
    \begin{tabular}{@{} p{0.085\textwidth} p{0.88\textwidth} @{}}
        \toprule
        Activity & Description and example \\
        \midrule
        
        Adapt & Modify a solution for other domains/ranges, i.e., change a solution to fit in a given context. Example: Remove preprocessor directives to reduce variability in a family of systems~\citep{Schulze2013}. \\
        
        Analyze & Probe the complexity of a solution by reading the source code. Example: Identify the function where the program would spend more time running.\\
       
        Apply & Use an existing solution as a component to implement a part of a program. Example: Reuse of off-the-shelf components. \\
          
        Debug & Detect and correct flaws in a design. Example: Given a program, identify faults and fix them~\citep{Scanniello2013}. \\
        
        Design & Understand a program and devise a solution for a new problem specification. Example: Given a program, sketch a solution for a new functionality. \\
        
        Implement & Read (part of) a program and implement a solution given a completed design. Example: Write code using the given examples according to a specification~\citep{Stefik2013}. \\
        
        Model & Illustrate or create an abstraction of a solution by reading the source code. Example: Given a program and the design of a new functionality, construct a UML model representing it. \\
       
        Present & Explain a solution to others. Example: Read a program and then write a description of what it does and how~\citep{Chaudhary1980}. \\
        
        Recognize & Identify a concept or code structure from a base knowledge or vocabulary of the domain obtained before the task to be performed. Example: ``The sorting algorithm (lines 46-62) can best be described as: (A) bubble sort (B) selection sort (C) heap sort (D) string sort (E) partition exchange sort''~\citep{Tenny1988}. \\
       
        Refactor & Redesign a solution to modify non-functional properties of a program or, at a larger scale, re-engineer it. Example: Rewrite a function so as to avoid using conditional expressions.  \\
       
        Relate & Understand different solutions and identify distinctions and similarities, pros and cons of these solutions. Example: Choose one out of three high-level descriptions that best describe the function of a previously studied application~\citep{Blinman2005}. \\
        
        Trace & Desk-check a solution, i.e., simulate program execution while looking at its code. Example: Consider the fragment ``x=++y'': what is the final value of x if y is -10?~\citep{Dolado2003}. \\
        
        Inspect* & Examine a source code to find or understand fine-grain static elements. Inspect is similar to Analyze, but the inspected aspect of the program is static instead of dynamic. Example: ``All variables in this program are global [true/false] '' \citep{Miara1983}. \\

        Memorize* & Memorize a source code in order to reconstruct it later, partially or as a whole. Example: Given a program, memorize it in 3 minutes and then reconstruct it in 4~\citep{Love1977}. \\
        
        \bottomrule
    \end{tabular}
\end{table*}

%% file: tables/tasks-activities-mapping.tex
\rowcolors{2}{black!15}{white}
\setlength{\tabcolsep}{3.4pt} 

\begin{table*}[t]
    \caption{Mapping of tasks (rows) to learning activities (columns), adapted from~\cite{oliveira2020}. The rows exhibit tasks that subjects performed during the investigated studies and the columns present the activities explained in \autoref{tab:learning-activities}. Opinion is market with ``*'' because it is not directly related to learning.}
    \label{tab:tasks-activities-mapping}
    \centering
    \scriptsize
    \begin{tabular}{lcccccccccc}
        \toprule
        {} & \rotatebox{60}{\parbox{1cm}{Adapt}} & \rotatebox{60}{\parbox{1cm}{Debug}} & \rotatebox{60}{\parbox{1cm}{Implement}} & \rotatebox{60}{\parbox{1cm}{Inspect}} & \rotatebox{60}{\parbox{1cm}{Memorize}} & \rotatebox{60}{\parbox{1cm}{Present}} & \rotatebox{60}{\parbox{1cm}{Recognize}} & \rotatebox{60}{\parbox{1cm}{Relate}} & \rotatebox{60}{\parbox{1cm}{Trace}} & \rotatebox{60}{\parbox{1.3cm}{Opinion*}} \\
        \midrule
        
        \multicolumn{11}{@{}l}{\textbf{\underline{provide information about the code}}} \\
        explain what the code does & & & & & & \cmark & & \cmark & & \\
        answer questions about code & & & & \cmark & & & \cmark & \cmark & \cmark & \\
        remember (part of) the code & & & \cmark & & \cmark & \cmark & & & & \\
        
        \multicolumn{11}{@{}l}{\textbf{\underline{act on the code}}} \\
        find and fix bugs in the code & & \cmark & & & & & & & & \\
        modify the code & \cmark & & \cmark & \cmark & & & & & \cmark & \\
        write code & & & \cmark & \cmark & & & & & & \\
        
        \multicolumn{11}{@{}l}{\textbf{\underline{provide personal opinion}}} \\
        opinion about the code & & & & & & & & & & \cmark \\
        answer if understood the code & & & & & & & & & & \cmark \\
        rate confidence in her answer & & & & & & \cmark & & & \cmark & \cmark \\
        rate the task difficulty & & \cmark & \cmark & \cmark & & \cmark & & & \cmark & \cmark \\
        
        \bottomrule
    \end{tabular}
\end{table*}

\rowcolors{2}{}{}

%% file: tables/card-results.tex
\begin{table}[t]
    \caption{Card sorting results.}
    \label{tab:card-results}
    \centering
    \scriptsize
    \tabulinesep=2pt
    \begin{tabu}{@{}p{5.5cm} p{7.9cm}@{}}
        \toprule
        Groups & Formatting elements (factors) \\
        \midrule
        
        \multirow{2}{*}{\hyperref[sec:formatting_styles]{Formatting}} & Formatting style \\
        & Formatting layout \\
        
        \tabucline[0.4pt gray!80 off 5pt]{-}
        
        \multirow{5}{*}{\hyperref[sec:spacing]{Spacing}} & Indentation \\
        & Appropriate use of indentation with blocks \\
        & Vertical and horizontal spacing \\
        & Vertical spacing between related instructions \\
        & Blank space around operators and parameters \\
        
        \tabucline[0.4pt gray!80 off 5pt]{-}
        
        \multirow{3}{*}{\hyperref[sec:block_delimiters]{Block Delimiters}} 
        & Block delimiter location \\
        & Block delimiter style \\
        & Block delimiter visibility \\
        
        \tabucline[0.4pt gray!80 off 5pt]{-}
        
        \multirow{2}{*}{\hyperref[sec:long_or_complex_code_line]{Long or Complex Code Line}} & Line length \\
        & Statements per line \\
        
        \tabucline[0.4pt gray!80 off 5pt]{-}
        
        \hyperref[sec:word_boundary_styles]{Word Boundary Styles} & Identifier style \\
        
        \bottomrule
    \end{tabu}
\end{table}

%% file: tables/included-papers.tex
\begin{table}[t!]
    \caption{Included papers and the factors analyzed, sorted in descending order by year of publication.}
    \label{tab:included-papers}
    \centering
    \scriptsize
    \begin{tabular}{p{2.4cm} p{4.9cm} l p{4.2cm}}
        \toprule
        Study & Study title & PL & Formatting elements (factors) \\
        \midrule
        
        \rowcolor{black!15} \cite{langhout2021} & ``Atoms of Confusion in Java'' & Java & -- Appropriate use of indentation \newline \hspace*{1.5mm} with blocks \newline -- Block delimiter visibility \\
        \cite{Bauer2019} & ``Indentation: Simply a Matter of Style or Support for Program Comprehension?'' & Java & -- Indentation \\
        \rowcolor{black!15} \cite{Medeiros2019} & ``An investigation of misunderstanding code patterns in C open-source software projects'' & C & -- Block delimiter visibility \newline -- Statements per line \\
        \cite{Santos2018} & ``Impacts of Coding Practices on Readability'' & Java & -- Indentation \newline -- Vertical spacing between \newline \hspace*{1.5mm} related instructions \newline -- Block delimiter location \newline -- Line length \newline -- Statements per line \\
        \rowcolor{black!15} \cite{Siegmund2017} & ``Measuring Neural Efficiency of Program Comprehension'' & Java & -- Formatting layout \\
        \cite{Gopstein2017} & ``Understanding Misunderstandings in Source Code'' & C/C++ & -- Block delimiter visibility \\
        \rowcolor{black!15} \cite{sampaio2016} & ``Software readability practices and the importance of their teaching'' & Java & -- Blank space around operators \newline \hspace*{1.5mm} and parameters \newline -- Block delimiter visibility \newline -- Statements per line \\
        \cite{Binkley2013} & ``The impact of identifier style on effort and comprehension'' & C/Java & -- Identifier style \\
        \rowcolor{black!15} \cite{Sharif2010} & ``An Eye Tracking Study on camelCase and under\_score Identifier Styles'' & None & -- Identifier style \\
        \cite{Furman2002} & ``A Look at Programmers Communicating through Program Indentation'' & Pascal & -- Indentation \\
        \rowcolor{black!15} \cite{Arab1992} & ``Enhancing Program Comprehension: Formatting and Documenting'' & Pascal & -- Block delimiter location \\
        \cite{Oman1990} & ``Typographic Style is More than Cosmetic'' & Pascal/C & -- Formatting style \\
        \rowcolor{black!15} \cite{Miara1983} & ``Program Indentation and Comprehensibility'' & Pascal & -- Indentation \newline -- Block delimiter location \\
        \cite{Sykes1983} & ``The Effect of Scope Delimiters on Program Comprehension'' & Pascal & -- Block delimiter visibility \newline -- Block delimiter style \\
        \rowcolor{black!15} \cite{Love1977} & ``An Experimental Investigation of the Effect of Program Structure on Program Understanding'' & Pascal & -- Vertical and horizontal spacing \\

        \bottomrule
    \end{tabular}
\end{table}

%% file: tables/summary-formatting-styles.tex
\begin{table}[t]
    \caption{Summary of results for the Formatting group.}
    \label{summary-formatting-styles}
    \centering
    \scriptsize
    \begin{tabular}{@{}p{3.3cm}p{10.1cm}@{}}
        \toprule
        \multicolumn{2}{@{}l}{\colorbox{black!15}{Factor} -- \underline{Study: Levels (Programming Language)} -- Dependent Variables (Activities): Results} \\
        \midrule
        
        \rowcolor{black!15} \multicolumn{2}{l}{Formatting style} \\ 
        \multicolumn{2}{l}{\tabindent \underline{\cite{Oman1990}: book format and Lightspeed Pascal style (Pascal)}}\\ 
        & Correctness (Trace): The book format style is the best.\\
        & Time (Trace): No significant difference.\\
        & Opinion: The book format style is the best. \\
        
        \multicolumn{2}{l}{\tabindent \underline{\cite{Oman1990}: book format and Kernighan \& Ritchie style (C)}} \\
        & Correctness (Trace): The book format style is the best.\\
        & Time (Trace): No significant difference. \\
        & Opinion: No significant difference. \\ 
        
        \rowcolor{black!15} \multicolumn{2}{l}{Formatting layout}\\
        \multicolumn{2}{l}{\tabindent \underline{\cite{Siegmund2017}: pretty-printed and disrupted (Java)}} \\
        & Brain Metrics (Relate): No significant difference. \\
        
        \bottomrule
    \end{tabular}
\end{table}

%% file: tables/summary-spacing.tex
\begin{table}[t]
    \caption{Summary of results for the Spacing group.}
    \label{summary-spacing}
    \centering
    \scriptsize
    \begin{tabular}{@{}p{3.3cm}p{10.1cm}@{}}
        \toprule
        \multicolumn{2}{@{}l}{\colorbox{black!15}{Factor} -- \underline{Study: Levels (Programming Language)} -- Dependent Variables (Activities): Results} \\
        \midrule
        
        \rowcolor{black!15} \multicolumn{2}{l}{Indentation} \\ 
        \multicolumn{2}{l}{\tabindent \underline{\cite{Miara1983}: 0, 2, 4, and 6 (Pascal)}} \\
        & Correctness (Trace, Inspect, Present): Two-space indentation is the best. \\
        & Opinion: Two-space indentation is the best. \\

        \multicolumn{2}{l}{\tabindent \underline{\cite{Furman2002}: left [0], normal [2-4], and random (Pascal)}} \\
        & Correctness (Trace, Inspect, Present): No significant difference. \\
        & Visual Metrics (Trace, Inspect, Present): Left and normal are better than random indentation on line-look time, normal is better than the other levels on revealed lines, and there is no difference on line-search time. \\
        & Opinion: Normal is considered less difficult than other indentation levels for subjective difficulty and preference. However, for fatigue, normal is considered only less difficult than random indentation. \\

        \multicolumn{2}{l}{\tabindent \underline{\cite{Santos2018}: 2 and 4 (Java)}} \\
        & Opinion: No significant difference. \\
        
        \multicolumn{2}{l}{\tabindent \underline{\cite{Bauer2019}: 0, 2, 4, and 8 (Java)}} \\ 
        & Correctness (Trace): No significant difference. \\
        & Time (Trace): No significant difference. \\
        & Visual Metrics (Trace): No significant difference. \\
        & Opinion: No significant difference. \\
        
        \rowcolor{black!15} \multicolumn{2}{l}{Appropriate use of indentation with blocks}\\
        \multicolumn{2}{l}{\tabindent \underline{\cite{langhout2021}: with and without (Java)}} \\
        & Correctness (Trace): Appropriate use of indentation is the best only when it is not preceded by curly braces. \\
        & Opinion: Appropriate use of indentation is the best. \\
        
        \rowcolor{black!15} \multicolumn{2}{l}{Vertical and horizontal spacing}\\
        
        \multicolumn{2}{l}{\tabindent \underline{\cite{Love1977}: paragraphed and unparagraphed (Pascal)}} \\
        & Correctness (Memorize and Present): No significant difference. \\
        
        \rowcolor{black!15} \multicolumn{2}{l}{Vertical spacing between related instructions}\\
        
        \multicolumn{2}{l}{\tabindent \underline{\cite{Santos2018}: with and without (Java)}} \\
        & Opinion: No significant difference. \\
        
        \rowcolor{black!15} \multicolumn{2}{l}{Blank space around operators and parameters}\\
        
        \multicolumn{2}{l}{\tabindent \underline{\cite{sampaio2016}: with and without (Java)}} \\
        & Opinion: No significant difference. \\
        
        \bottomrule
    \end{tabular}
\end{table}

%% file: tables/summary-block-delimiters.tex
\begin{table}[t]
    \caption{Summary of results for the Block Delimiters group.}
    \label{summary-block-delimiters}
    \centering
    \scriptsize
    \begin{tabular}{@{}p{3.3cm}p{10.1cm}@{}}
        \toprule
        \multicolumn{2}{@{}l}{\colorbox{black!15}{Factor} -- \underline{Study: Levels (Programming Language)} -- Dependent Variables (Activities): Results} \\
        \midrule
        
        \rowcolor{black!15} \multicolumn{2}{l}{Block delimiter location} \\
        \multicolumn{2}{l}{\tabindent \underline{\cite{Miara1983}: blocked and non-blocked (Pascal)}}\\
        &  Correctness (Trace, Inspect, Present): No significant difference. \\
        & Opinion: No significant difference. \\
        
        \multicolumn{2}{l}{\tabindent \underline{\cite{Arab1992}: in the statement line and in separate line (Pascal)}} \\
        & Opinion: Block delimiters in their own lines is the best.\\
        
        \multicolumn{2}{l}{\tabindent \underline{\cite{Santos2018}: in the statement line and in separate line (Java)}} \\
        & Opinion: Block delimiter in its own line is the best.\\

        \rowcolor{black!15} \multicolumn{2}{l}{Block delimiter style} \\
        \multicolumn{2}{l}{\tabindent \underline{\cite{Sykes1983}: \texttt{ENDIF/ENDWHILE} (without \texttt{BEGIN}) and \texttt{BEGIN-END} in all blocks (Pascal)}}\\
        & Correctness (Trace): \texttt{ENDIF/ENDWHILE} is the best.\\
        & Opinion: \texttt{ENDIF/ENDWHILE} is the best. \\
        
        \multicolumn{2}{l}{\tabindent \underline{\cite{Sykes1983}: \texttt{ENDIF/ENDWHILE} (without \texttt{BEGIN}) and \texttt{BEGIN-END} only in compound statement}}\\
        \multicolumn{2}{l}{\hspace{29.3mm}\underline{blocks (Pascal)}} \\
        & Correctness (Trace): \texttt{ENDIF/ENDWHILE} is the best. \\
        & Opinion: \texttt{ENDIF/ENDWHILE} is the best. \\
        
        \rowcolor{black!15} \multicolumn{2}{l}{Block delimiter visibility} \\ 
        \multicolumn{2}{l}{\tabindent \underline{\cite{Sykes1983}: omitted and present (Pascal)}}\\
        & Correctness (Trace): No significant difference.\\
        & Opinion: No significant difference.\\

        \multicolumn{2}{l}{\tabindent \underline{\cite{sampaio2016}: omitted and present (Java)}}\\
        & Opinion:  No significant difference. \\
        
        \multicolumn{2}{l}{\tabindent \underline{\cite{Gopstein2017}: omitted and present (C/C++)}}\\
        & Correctness (Trace): Present block delimiters is the best.\\
  
        \multicolumn{2}{l}{\tabindent \underline{\cite{Medeiros2019}: omitted and present (C)}}\\
        & Opinion: Present block delimiters is the best.\\

        \multicolumn{2}{l}{\tabindent \underline{\cite{langhout2021}: omitted and present (Java)}}\\
        & Correctness (Trace): Present block delimiters is the best.\\
        
    	\bottomrule
    \end{tabular}
\end{table}

%% file: tables/summary-long-complex-line.tex
\begin{table}[t]
    \caption{Summary of results for the Long or Complex Code Line group.}
    \label{summary-long-complex-line}
    \centering
    \scriptsize
    \begin{tabular}{@{}p{3.3cm}p{10.1cm}@{}}
        \toprule
        \multicolumn{2}{@{}l}{\colorbox{black!15}{Factor} -- \underline{Study: Levels (Programming Language)} -- Dependent Variables (Activities): Results} \\
        \midrule
        
        \rowcolor{black!15} \multicolumn{2}{l}{Line length} \\ 
        \multicolumn{2}{l}{\tabindent \underline{\cite{Santos2018}: limit of 80 characters and exceed of 80 characters (Java)}} \\
        & Opinion: Line length within the limit of 80 characters is the best.\\
        
        \rowcolor{black!15} \multicolumn{2}{l}{Statements per line}\\
        \multicolumn{2}{l}{\tabindent \underline{\cite{sampaio2016}: multiple and one (Java)}} \\
        & Opinion: One statement per line is relevant for code legibility. \\
        \multicolumn{2}{l}{\tabindent \underline{\cite{Santos2018}: multiple and one (Java)}} \\
        & Opinion: One statement per line is the best. \\ 
        \multicolumn{2}{l}{\tabindent \underline{\cite{Medeiros2019}: multiple and one (C)}} \\
        & Opinion: No significant difference. \\ 
        
        \bottomrule
    \end{tabular}
\end{table}

%% file: tables/summary-word-boundary-styles.tex
\begin{table}
    \caption{Summary of results for the Word Boundary Styles group.}
    \label{summary-word-boundary-styles}
    \centering
    \scriptsize
    \begin{tabular}{@{}p{3.3cm}p{10.1cm}@{}}
        \toprule
        \multicolumn{2}{@{}l}{\colorbox{black!15}{Factor} -- \underline{Study: Levels (Programming Language)} -- Dependent Variables (Activities): Results} \\
        \midrule
        
        \rowcolor{black!15} \multicolumn{2}{l}{Identifier style} \\ 
        \multicolumn{2}{l}{\tabindent \underline{\cite{Sharif2010}: camel case and underscore (None)}} \\ 
        & Correctness (Memorize): No significant difference.\\ 
        & Time (Memorize): Underscore is the best.\\
        & Visual Metrics (Memorize): Underscore is the best for average fixation duration.\\
        
        \multicolumn{2}{l}{\tabindent \underline{\cite{Binkley2013}: camel case and underscore (C and Java)}} \\
        & Correctness (Inspect, Memorize, Present): Camel case is the best only in the Inspect activity. \\
        & Time (Inspect, Present): Camel case is the best only in the Inspect activity. \\
        & Visual Metrics (Memorize): Camel case is the best. \\
        
        \bottomrule
    \end{tabular}
\end{table}

%% file: main.bbl
\begin{thebibliography}{56}
\expandafter\ifx\csname natexlab\endcsname\relax\def\natexlab#1{#1}\fi
\providecommand{\url}[1]{\texttt{#1}}
\providecommand{\href}[2]{#2}
\providecommand{\path}[1]{#1}
\providecommand{\DOIprefix}{doi:}
\providecommand{\ArXivprefix}{arXiv:}
\providecommand{\URLprefix}{URL: }
\providecommand{\Pubmedprefix}{pmid:}
\providecommand{\doi}[1]{\href{http://dx.doi.org/#1}{\path{#1}}}
\providecommand{\Pubmed}[1]{\href{pmid:#1}{\path{#1}}}
\providecommand{\bibinfo}[2]{#2}
\ifx\xfnm\relax \def\xfnm[#1]{\unskip,\space#1}\fi
\bibitem[{Smit et~al.(2011)Smit, Gergel, Hoover, and Stroulia}]{smit2011}
\bibinfo{author}{M.~Smit}, \bibinfo{author}{B.~Gergel}, \bibinfo{author}{H.~J.
  Hoover}, \bibinfo{author}{E.~Stroulia},
\newblock \bibinfo{title}{{Code Convention Adherence in Evolving Software}},
\newblock in: \bibinfo{booktitle}{Proceedings of the 27th IEEE International
  Conference on Software Maintenance (ICSM '11)}, \bibinfo{publisher}{IEEE
  Computer Society}, \bibinfo{address}{USA}, \bibinfo{year}{2011}, p.
  \bibinfo{pages}{504–507}. \DOIprefix\doi{10.1109/ICSM.2011.6080819}.
\bibitem[{Miara et~al.(1983)Miara, Musselman, Navarro, and
  Shneiderman}]{Miara1983}
\bibinfo{author}{R.~J. Miara}, \bibinfo{author}{J.~A. Musselman},
  \bibinfo{author}{J.~A. Navarro}, \bibinfo{author}{B.~Shneiderman},
\newblock \bibinfo{title}{{Program Indentation and Comprehensibility}},
\newblock \bibinfo{journal}{Communications of the ACM} \bibinfo{volume}{26}
  (\bibinfo{year}{1983}) \bibinfo{pages}{861--867}.
  \DOIprefix\doi{10.1145/182.358437}.
\bibitem[{Binkley et~al.(2013)Binkley, Davis, Lawrie, Maletic, Morrell, and
  Sharif}]{Binkley2013}
\bibinfo{author}{D.~Binkley}, \bibinfo{author}{M.~Davis},
  \bibinfo{author}{D.~Lawrie}, \bibinfo{author}{J.~I. Maletic},
  \bibinfo{author}{C.~Morrell}, \bibinfo{author}{B.~Sharif},
\newblock \bibinfo{title}{{The impact of identifier style on effort and
  comprehension}},
\newblock \bibinfo{journal}{Empirical Software Engineering}
  \bibinfo{volume}{18} (\bibinfo{year}{2013}) \bibinfo{pages}{219--276}.
  \DOIprefix\doi{10.1007/s10664-012-9201-4}.
\bibitem[{Oliveira et~al.(2020)Oliveira, Bruno, Madeiral, and
  Castor}]{oliveira2020}
\bibinfo{author}{D.~Oliveira}, \bibinfo{author}{R.~Bruno},
  \bibinfo{author}{F.~Madeiral}, \bibinfo{author}{F.~Castor},
\newblock \bibinfo{title}{{Evaluating Code Readability and Legibility: An
  Examination of Human-centric Studies}},
\newblock in: \bibinfo{booktitle}{Proceedings of the 36th IEEE International
  Conference on Software Maintenance and Evolution (ICSME '20)},
  \bibinfo{organization}{IEEE}, \bibinfo{year}{2020}, pp.
  \bibinfo{pages}{348--359}. \DOIprefix\doi{10.1109/ICSME46990.2020.00041}.
\bibitem[{Santos and Gerosa(2018)}]{Santos2018}
\bibinfo{author}{R.~M.~d. Santos}, \bibinfo{author}{M.~A. Gerosa},
\newblock \bibinfo{title}{{Impacts of Coding Practices on Readability}},
\newblock in: \bibinfo{booktitle}{Proceedings of the 26th Conference on Program
  Comprehension (ICPC '18)}, \bibinfo{publisher}{ACM}, \bibinfo{address}{New
  York, NY, USA}, \bibinfo{year}{2018}, pp. \bibinfo{pages}{277--285}.
  \DOIprefix\doi{10.1145/3196321.3196342}.
\bibitem[{Bauer et~al.(2019)Bauer, Siegmund, Peitek, Hofmeister, and
  Apel}]{Bauer2019}
\bibinfo{author}{J.~Bauer}, \bibinfo{author}{J.~Siegmund},
  \bibinfo{author}{N.~Peitek}, \bibinfo{author}{J.~C. Hofmeister},
  \bibinfo{author}{S.~Apel},
\newblock \bibinfo{title}{{Indentation: Simply a Matter of Style or Support for
  Program Comprehension?}},
\newblock in: \bibinfo{booktitle}{Proceedings of the 27th International
  Conference on Program Comprehension (ICPC '19)}, \bibinfo{publisher}{IEEE
  Press}, \bibinfo{address}{Piscataway, NJ, USA}, \bibinfo{year}{2019}, pp.
  \bibinfo{pages}{154--164}. \DOIprefix\doi{10.1109/ICPC.2019.00033}.
\bibitem[{Sharif and Maletic(2010)}]{Sharif2010}
\bibinfo{author}{B.~Sharif}, \bibinfo{author}{J.~I. Maletic},
\newblock \bibinfo{title}{{An Eye Tracking Study on camelCase and under\_score
  Identifier Styles}},
\newblock in: \bibinfo{booktitle}{Proceedings of the 18th IEEE International
  Conference on Program Comprehension (ICPC '10)}, \bibinfo{publisher}{IEEE
  Computer Society}, \bibinfo{address}{Washington, DC, USA},
  \bibinfo{year}{2010}, pp. \bibinfo{pages}{196--205}.
  \DOIprefix\doi{10.1109/ICPC.2010.41}.
\bibitem[{Sampaio and Barbosa(2016)}]{sampaio2016}
\bibinfo{author}{I.~B. Sampaio}, \bibinfo{author}{L.~Barbosa},
\newblock \bibinfo{title}{Software readability practices and the importance of
  their teaching},
\newblock in: \bibinfo{booktitle}{Proceedings of the 7th International
  Conference on Information and Communication Systems (ICICS '16)},
  \bibinfo{organization}{IEEE}, \bibinfo{year}{2016}, pp.
  \bibinfo{pages}{304--309}. \DOIprefix\doi{10.1109/IACS.2016.7476069}.
\bibitem[{Benander et~al.(1996)Benander, Benander, and Pu}]{Benander1996}
\bibinfo{author}{A.~C. Benander}, \bibinfo{author}{B.~A. Benander},
  \bibinfo{author}{H.~Pu},
\newblock \bibinfo{title}{{Recursion vs. Iteration: An Empirical Study of
  Comprehension}},
\newblock \bibinfo{journal}{Journal of Systems and Software}
  \bibinfo{volume}{32} (\bibinfo{year}{1996}) \bibinfo{pages}{73--82}.
  \DOIprefix\doi{10.1016/0164-1212(95)00043-7}.
\bibitem[{Hofmeister et~al.(2019)Hofmeister, Siegmund, and
  Holt}]{Hofmeister2019}
\bibinfo{author}{J.~C. Hofmeister}, \bibinfo{author}{J.~Siegmund},
  \bibinfo{author}{D.~V. Holt},
\newblock \bibinfo{title}{Shorter identifier names take longer to comprehend},
\newblock \bibinfo{journal}{Empirical Software Engineering}
  \bibinfo{volume}{24} (\bibinfo{year}{2019}) \bibinfo{pages}{417--443}.
  \DOIprefix\doi{10.1007/s10664-018-9621-x}.
\bibitem[{Bloom et~al.(1956)Bloom, Engelhart, Furst, Hill, and
  Krathwohl}]{bloom1956taxonomy}
\bibinfo{author}{B.~Bloom}, \bibinfo{author}{M.~Engelhart},
  \bibinfo{author}{E.~Furst}, \bibinfo{author}{W.~H. Hill},
  \bibinfo{author}{D.~R. Krathwohl}, \bibinfo{title}{{Taxonomy of Educational
  Objectives: The Classification of Educational Goals. Handbook I: Cognitive
  Domain}}, \bibinfo{publisher}{David McKay Company}, \bibinfo{address}{New
  York}, \bibinfo{year}{1956}.
\bibitem[{Fuller et~al.(2007)Fuller, Johnson, Ahoniemi, Cukierman,
  Hern\'{a}n-Losada, Jackova, Lahtinen, Lewis, Thompson, Riedesel, and
  Thompson}]{fuller2007developing}
\bibinfo{author}{U.~Fuller}, \bibinfo{author}{C.~G. Johnson},
  \bibinfo{author}{T.~Ahoniemi}, \bibinfo{author}{D.~Cukierman},
  \bibinfo{author}{I.~Hern\'{a}n-Losada}, \bibinfo{author}{J.~Jackova},
  \bibinfo{author}{E.~Lahtinen}, \bibinfo{author}{T.~L. Lewis},
  \bibinfo{author}{D.~M. Thompson}, \bibinfo{author}{C.~Riedesel},
  \bibinfo{author}{E.~Thompson},
\newblock \bibinfo{title}{{Developing a Computer Science-specific Learning
  Taxonomy}},
\newblock \bibinfo{journal}{ACM SIGCSE Bulletin} \bibinfo{volume}{39}
  (\bibinfo{year}{2007}) \bibinfo{pages}{152--170}.
  \DOIprefix\doi{10.1145/1345375.1345438}.
\bibitem[{Buse and Weimer(2010)}]{Buse2010}
\bibinfo{author}{R.~P.~L. Buse}, \bibinfo{author}{W.~R. Weimer},
\newblock \bibinfo{title}{{Learning a Metric for Code Readability}},
\newblock \bibinfo{journal}{IEEE Transactions on Software Engineering}
  \bibinfo{volume}{36} (\bibinfo{year}{2010}) \bibinfo{pages}{546--558}.
  \DOIprefix\doi{10.1109/TSE.2009.70}.
\bibitem[{Almeida et~al.(2003)Almeida, Camargo, Basseto, and Paz}]{de2003best}
\bibinfo{author}{J.~R.~d. Almeida}, \bibinfo{author}{J.~B. Camargo},
  \bibinfo{author}{B.~A. Basseto}, \bibinfo{author}{S.~M. Paz},
\newblock \bibinfo{title}{{Best Practices in Code Inspection for
  Safety-Critical Software}},
\newblock \bibinfo{journal}{IEEE Software} \bibinfo{volume}{20}
  (\bibinfo{year}{2003}) \bibinfo{pages}{56--63}.
  \DOIprefix\doi{10.1109/MS.2003.1196322}.
\bibitem[{Gough and Tunmer(1986)}]{gough1986decoding}
\bibinfo{author}{P.~B. Gough}, \bibinfo{author}{W.~E. Tunmer},
\newblock \bibinfo{title}{{Decoding, Reading, and Reading Disability}},
\newblock \bibinfo{journal}{Remedial and Special Education} \bibinfo{volume}{7}
  (\bibinfo{year}{1986}) \bibinfo{pages}{6--10}.
  \DOIprefix\doi{10.1177/074193258600700104}.
\bibitem[{Hoover and Gough(1990)}]{hoover1990simple}
\bibinfo{author}{W.~A. Hoover}, \bibinfo{author}{P.~B. Gough},
\newblock \bibinfo{title}{The simple view of reading},
\newblock \bibinfo{journal}{Reading and Writing} \bibinfo{volume}{2}
  (\bibinfo{year}{1990}) \bibinfo{pages}{127--160}.
  \DOIprefix\doi{10.1007/BF00401799}.
\bibitem[{DuBay(2004)}]{dubay2004principles}
\bibinfo{author}{W.~H. DuBay},
\newblock \bibinfo{title}{{The Principles of Readability}},
\newblock \bibinfo{journal}{Online Submission}  (\bibinfo{year}{2004}).
\bibitem[{Tekfi(1987)}]{tekfi1987readability}
\bibinfo{author}{C.~Tekfi},
\newblock \bibinfo{title}{{Readability Formulas: An Overview}},
\newblock \bibinfo{journal}{Journal of Documentation} \bibinfo{volume}{43}
  (\bibinfo{year}{1987}) \bibinfo{pages}{261--273}.
  \DOIprefix\doi{10.1108/eb026811}.
\bibitem[{Daka et~al.(2015)Daka, Campos, Fraser, Dorn, and Weimer}]{daka2015}
\bibinfo{author}{E.~Daka}, \bibinfo{author}{J.~Campos},
  \bibinfo{author}{G.~Fraser}, \bibinfo{author}{J.~Dorn},
  \bibinfo{author}{W.~Weimer},
\newblock \bibinfo{title}{{Modeling Readability to Improve Unit Tests}},
\newblock in: \bibinfo{booktitle}{Proceedings of the 10th Joint Meeting on
  Foundations of Software Engineering (ESEC/FSE '15)},
  \bibinfo{publisher}{Association for Computing Machinery},
  \bibinfo{address}{New York, NY, USA}, \bibinfo{year}{2015}, pp.
  \bibinfo{pages}{107--118}. \DOIprefix\doi{10.1145/2786805.2786838}.
\bibitem[{Strizver(2013)}]{strizver2013type}
\bibinfo{author}{I.~Strizver}, \bibinfo{title}{{Type Rules: The designer's
  guide to professional typography}}, \bibinfo{publisher}{John Wiley \& Sons},
  \bibinfo{year}{2013}.
\bibitem[{Zuffi et~al.(2007)Zuffi, Brambilla, Beretta, and
  Scala}]{zuffi2007human}
\bibinfo{author}{S.~Zuffi}, \bibinfo{author}{C.~Brambilla},
  \bibinfo{author}{G.~Beretta}, \bibinfo{author}{P.~Scala},
\newblock \bibinfo{title}{{Human Computer Interaction: Legibility and
  Contrast}},
\newblock in: \bibinfo{booktitle}{Proceedings of the 14th International
  Conference on Image Analysis and Processing (ICIAP '07)},
  \bibinfo{organization}{IEEE}, \bibinfo{year}{2007}, pp.
  \bibinfo{pages}{241--246}. \DOIprefix\doi{10.1109/ICIAP.2007.4362786}.
\bibitem[{Feitelson(2022)}]{feitelson2022}
\bibinfo{author}{D.~G. Feitelson},
\newblock \bibinfo{title}{Considerations and pitfalls for reducing threats to
  the validity of controlled experiments on code comprehension},
\newblock \bibinfo{journal}{Empirical Software Engineering}
  \bibinfo{volume}{27} (\bibinfo{year}{2022}).
  \DOIprefix\doi{10.1007/s10664-022-10160-3}.
\bibitem[{Gopstein et~al.(2017)Gopstein, Iannacone, Yan, DeLong, Zhuang, Yeh,
  and Cappos}]{Gopstein2017}
\bibinfo{author}{D.~Gopstein}, \bibinfo{author}{J.~Iannacone},
  \bibinfo{author}{Y.~Yan}, \bibinfo{author}{L.~DeLong},
  \bibinfo{author}{Y.~Zhuang}, \bibinfo{author}{M.~K.-C. Yeh},
  \bibinfo{author}{J.~Cappos},
\newblock \bibinfo{title}{{Understanding Misunderstandings in Source Code}},
\newblock in: \bibinfo{booktitle}{Proceedings of the 11th Joint Meeting on
  Foundations of Software Engineering (ESEC/FSE '17)},
  \bibinfo{publisher}{ACM}, \bibinfo{address}{New York, NY, USA},
  \bibinfo{year}{2017}, pp. \bibinfo{pages}{129--139}.
  \DOIprefix\doi{10.1145/3106237.3106264}.
\bibitem[{Ajami et~al.(2019)Ajami, Woodbridge, and Feitelson}]{Ajami2019}
\bibinfo{author}{S.~Ajami}, \bibinfo{author}{Y.~Woodbridge},
  \bibinfo{author}{D.~G. Feitelson},
\newblock \bibinfo{title}{{Syntax, predicates, idioms -- what really affects
  code complexity?}},
\newblock \bibinfo{journal}{Empirical Software Engineering}
  \bibinfo{volume}{24} (\bibinfo{year}{2019}) \bibinfo{pages}{287--328}.
  \DOIprefix\doi{10.1007/s10664-018-9628-3}.
\bibitem[{Scanniello and Risi(2013)}]{Scanniello2013}
\bibinfo{author}{G.~Scanniello}, \bibinfo{author}{M.~Risi},
\newblock \bibinfo{title}{{Dealing with Faults in Source Code: Abbreviated vs.
  Full-Word Identifier Names}},
\newblock in: \bibinfo{booktitle}{Proceedings of the 2013 IEEE International
  Conference on Software Maintenance (ICSM '13)}, \bibinfo{publisher}{IEEE
  Computer Society}, \bibinfo{address}{USA}, \bibinfo{year}{2013}, pp.
  \bibinfo{pages}{190--199}. \DOIprefix\doi{10.1109/ICSM.2013.30}.
\bibitem[{Jbara and Feitelson(2014)}]{Jbara2014b}
\bibinfo{author}{A.~Jbara}, \bibinfo{author}{D.~G. Feitelson},
\newblock \bibinfo{title}{{On the Effect of Code Regularity on Comprehension}},
\newblock in: \bibinfo{booktitle}{Proceedings of the 22nd International
  Conference on Program Comprehension (ICPC '14)}, \bibinfo{publisher}{ACM},
  \bibinfo{address}{New York, NY, USA}, \bibinfo{year}{2014}, pp.
  \bibinfo{pages}{189--200}. \DOIprefix\doi{10.1145/2597008.2597140}.
\bibitem[{O'Neal and Edwards(1994)}]{ONeal1994}
\bibinfo{author}{M.~B. O'Neal}, \bibinfo{author}{W.~R. Edwards},
\newblock \bibinfo{title}{{Complexity Measures for Rule-Based Programs}},
\newblock \bibinfo{journal}{IEEE Transactions on Knowledge and Data
  Engineering} \bibinfo{volume}{6} (\bibinfo{year}{1994})
  \bibinfo{pages}{669--680}. \DOIprefix\doi{10.1109/69.317699}.
\bibitem[{Siegmund et~al.(2017)Siegmund, Peitek, Parnin, Apel, Hofmeister,
  K\"{a}stner, Begel, Bethmann, and Brechmann}]{Siegmund2017}
\bibinfo{author}{J.~Siegmund}, \bibinfo{author}{N.~Peitek},
  \bibinfo{author}{C.~Parnin}, \bibinfo{author}{S.~Apel},
  \bibinfo{author}{J.~Hofmeister}, \bibinfo{author}{C.~K\"{a}stner},
  \bibinfo{author}{A.~Begel}, \bibinfo{author}{A.~Bethmann},
  \bibinfo{author}{A.~Brechmann},
\newblock \bibinfo{title}{{Measuring Neural Efficiency of Program
  Comprehension}},
\newblock in: \bibinfo{booktitle}{Proceedings of the 11th Joint Meeting on
  Foundations of Software Engineering (ESEC/FSE '17)},
  \bibinfo{publisher}{ACM}, \bibinfo{address}{New York, NY, USA},
  \bibinfo{year}{2017}, pp. \bibinfo{pages}{140--150}.
  \DOIprefix\doi{10.1145/3106237.3106268}.
\bibitem[{Anderson et~al.(2001)Anderson, Bloom, Krathwohl, Airasian,
  Cruikshank, Mayer, Pintrich, Raths, and Wittrock}]{bloomrevised2001}
\bibinfo{author}{L.~Anderson}, \bibinfo{author}{B.~Bloom},
  \bibinfo{author}{D.~Krathwohl}, \bibinfo{author}{P.~Airasian},
  \bibinfo{author}{K.~Cruikshank}, \bibinfo{author}{R.~Mayer},
  \bibinfo{author}{P.~Pintrich}, \bibinfo{author}{J.~Raths},
  \bibinfo{author}{M.~Wittrock}, \bibinfo{title}{{A Taxonomy for Learning,
  Teaching, and Assessing: A Revision of Bloom's Taxonomy of Educational
  Objectives}}, \bibinfo{publisher}{Longman}, \bibinfo{year}{2001}.
\bibitem[{Schulze et~al.(2013)Schulze, Liebig, Siegmund, and
  Apel}]{Schulze2013}
\bibinfo{author}{S.~Schulze}, \bibinfo{author}{J.~Liebig},
  \bibinfo{author}{J.~Siegmund}, \bibinfo{author}{S.~Apel},
\newblock \bibinfo{title}{{Does the Discipline of Preprocessor Annotations
  Matter? A Controlled Experiment}},
\newblock in: \bibinfo{booktitle}{Proceedings of the 12th International
  Conference on Generative Programming: Concepts \& Experiences (GPCE '13)},
  \bibinfo{publisher}{ACM}, \bibinfo{address}{New York, NY, USA},
  \bibinfo{year}{2013}, pp. \bibinfo{pages}{65--74}.
  \DOIprefix\doi{10.1145/2517208.2517215}.
\bibitem[{Stefik and Siebert(2013)}]{Stefik2013}
\bibinfo{author}{A.~Stefik}, \bibinfo{author}{S.~Siebert},
\newblock \bibinfo{title}{{An Empirical Investigation into Programming Language
  Syntax}},
\newblock \bibinfo{journal}{ACM Transactions on Computing Education}
  \bibinfo{volume}{13} (\bibinfo{year}{2013}) \bibinfo{pages}{19:1--19:40}.
  \DOIprefix\doi{10.1145/2534973}.
\bibitem[{Chaudhary and Sahasrabuddhe(1980)}]{Chaudhary1980}
\bibinfo{author}{B.~D. Chaudhary}, \bibinfo{author}{H.~V. Sahasrabuddhe},
\newblock \bibinfo{title}{{Meaningfulness as a Factor of Program Complexity}},
\newblock in: \bibinfo{booktitle}{Proceedings of the ACM 1980 Annual Conference
  (ACM '80)}, \bibinfo{publisher}{ACM}, \bibinfo{address}{New York, NY, USA},
  \bibinfo{year}{1980}, pp. \bibinfo{pages}{457--466}.
  \DOIprefix\doi{10.1145/800176.810001}.
\bibitem[{{Tenny}(1988)}]{Tenny1988}
\bibinfo{author}{T.~{Tenny}},
\newblock \bibinfo{title}{{Program Readability: Procedures Versus Comments}},
\newblock \bibinfo{journal}{IEEE Transactions on Software Engineering}
  \bibinfo{volume}{14} (\bibinfo{year}{1988}) \bibinfo{pages}{1271--1279}.
  \DOIprefix\doi{10.1109/32.6171}.
\bibitem[{Blinman and Cockburn(2005)}]{Blinman2005}
\bibinfo{author}{S.~Blinman}, \bibinfo{author}{A.~Cockburn},
\newblock \bibinfo{title}{{Program Comprehension: Investigating the Effects of
  Naming Style and Documentation}},
\newblock in: \bibinfo{booktitle}{Proceedings of the 6th Australasian User
  Interface Conference (AUIC '05)}, \bibinfo{publisher}{ACS},
  \bibinfo{address}{Newcastle, Australia}, \bibinfo{year}{2005}, pp.
  \bibinfo{pages}{73--78}.
\bibitem[{Dolado et~al.(2003)Dolado, Harman, Otero, and Hu}]{Dolado2003}
\bibinfo{author}{J.~J. Dolado}, \bibinfo{author}{M.~Harman},
  \bibinfo{author}{M.~C. Otero}, \bibinfo{author}{L.~Hu},
\newblock \bibinfo{title}{{An Empirical Investigation of the Influence of a
  Type of Side Effects on Program Comprehension}},
\newblock \bibinfo{journal}{IEEE Transactions on Software Engineering}
  \bibinfo{volume}{29} (\bibinfo{year}{2003}) \bibinfo{pages}{665--670}.
  \DOIprefix\doi{10.1109/TSE.2003.1214329}.
\bibitem[{Love(1977)}]{Love1977}
\bibinfo{author}{T.~Love},
\newblock \bibinfo{title}{{An Experimental Investigation of the Effect of
  Program Structure on Program Understanding}},
\newblock \bibinfo{journal}{ACM SIGOPS Operating Systems Review -- Proceedings
  of an ACM Conference on Language Design for Reliable Software}
  \bibinfo{volume}{11} (\bibinfo{year}{1977}) \bibinfo{pages}{105--113}.
  \DOIprefix\doi{10.1145/390018.808317}.
\bibitem[{Kitchenham et~al.(2015)Kitchenham, Budgen, and
  Brereton}]{Kitchenham2015}
\bibinfo{author}{B.~A. Kitchenham}, \bibinfo{author}{D.~Budgen},
  \bibinfo{author}{P.~Brereton}, \bibinfo{title}{{Evidence-Based Software
  Engineering and Systematic Reviews}}, \bibinfo{publisher}{Chapman \&
  Hall/CRC}, \bibinfo{year}{2015}.
\bibitem[{Binkley et~al.(2009)Binkley, Davis, Lawrie, and
  Morrell}]{binkley2009b}
\bibinfo{author}{D.~W. Binkley}, \bibinfo{author}{M.~Davis},
  \bibinfo{author}{D.~J. Lawrie}, \bibinfo{author}{C.~Morrell},
\newblock \bibinfo{title}{{To CamelCase or Under\_score}},
\newblock in: \bibinfo{booktitle}{Proceedings of the 17th International
  Conference on Program Comprehension (ICPC '09)}, \bibinfo{publisher}{IEEE
  Press}, \bibinfo{address}{Piscataway, NJ, USA}, \bibinfo{year}{2009}, pp.
  \bibinfo{pages}{158--167}. \DOIprefix\doi{10.1109/ICPC.2009.5090039}.
\bibitem[{Keele et~al.(2007)}]{keele2007guidelines}
\bibinfo{author}{S.~Keele}, et~al., \bibinfo{title}{{Guidelines for Performing
  Systematic Literature Reviews in Software Engineering}},
  \bibinfo{type}{Technical Report}, Version 2.3 EBSE, \bibinfo{year}{2007}.
\bibitem[{Wood and Wood(2008)}]{wood2008card}
\bibinfo{author}{J.~R. Wood}, \bibinfo{author}{L.~E. Wood},
\newblock \bibinfo{title}{{Card Sorting: Current Practices and Beyond}},
\newblock \bibinfo{journal}{Journal of Usability Studies} \bibinfo{volume}{4}
  (\bibinfo{year}{2008}) \bibinfo{pages}{1–6}.
\bibitem[{Arab(1992)}]{Arab1992}
\bibinfo{author}{M.~Arab},
\newblock \bibinfo{title}{{Enhancing Program Comprehension: Formatting and
  Documenting}},
\newblock \bibinfo{journal}{ACM SIGPLAN Notices} \bibinfo{volume}{27}
  (\bibinfo{year}{1992}) \bibinfo{pages}{37--46}.
  \DOIprefix\doi{10.1145/130973.130975}.
\bibitem[{Langhout and Aniche(2021)}]{langhout2021}
\bibinfo{author}{C.~Langhout}, \bibinfo{author}{M.~Aniche},
\newblock \bibinfo{title}{{Atoms of Confusion in Java}},
\newblock in: \bibinfo{booktitle}{Proceedings of the 29th IEEE/ACM
  International Conference on Program Comprehension (ICPC '21)},
  \bibinfo{organization}{IEEE}, \bibinfo{year}{2021}, pp.
  \bibinfo{pages}{25--35}. \DOIprefix\doi{10.1109/ICPC52881.2021.00012}.
\bibitem[{Medeiros et~al.(2019)Medeiros, Lima, Amaral, Apel, K\"{a}stner,
  Ribeiro, and Gheyi}]{Medeiros2019}
\bibinfo{author}{F.~Medeiros}, \bibinfo{author}{G.~Lima},
  \bibinfo{author}{G.~Amaral}, \bibinfo{author}{S.~Apel},
  \bibinfo{author}{C.~K\"{a}stner}, \bibinfo{author}{M.~Ribeiro},
  \bibinfo{author}{R.~Gheyi},
\newblock \bibinfo{title}{{An investigation of misunderstanding code patterns
  in C open-source software projects}},
\newblock \bibinfo{journal}{Empirical Software Engineering}
  \bibinfo{volume}{24} (\bibinfo{year}{2019}) \bibinfo{pages}{1693--1726}.
  \DOIprefix\doi{10.1007/s10664-018-9666-x}.
\bibitem[{Furman et~al.(2002)Furman, Boehm-Davis, and Holt}]{Furman2002}
\bibinfo{author}{S.~Furman}, \bibinfo{author}{D.~A. Boehm-Davis},
  \bibinfo{author}{R.~W. Holt},
\newblock \bibinfo{title}{{A Look at Programmers Communicating through Program
  Indentation}},
\newblock \bibinfo{journal}{Journal of the Washington Academy of Sciences}
  \bibinfo{volume}{88} (\bibinfo{year}{2002}) \bibinfo{pages}{73--88}.
\bibitem[{Oman and Cook(1990)}]{Oman1990}
\bibinfo{author}{P.~W. Oman}, \bibinfo{author}{C.~R. Cook},
\newblock \bibinfo{title}{{Typographic Style is More than Cosmetic}},
\newblock \bibinfo{journal}{Communications of the ACM} \bibinfo{volume}{33}
  (\bibinfo{year}{1990}) \bibinfo{pages}{506--520}.
  \DOIprefix\doi{10.1145/78607.78611}.
\bibitem[{Sykes et~al.(1983)Sykes, Tillman, and Shneiderman}]{Sykes1983}
\bibinfo{author}{F.~Sykes}, \bibinfo{author}{R.~T. Tillman},
  \bibinfo{author}{B.~Shneiderman},
\newblock \bibinfo{title}{{The Effect of Scope Delimiters on Program
  Comprehension}},
\newblock \bibinfo{journal}{Software: Practice and Experience}
  \bibinfo{volume}{13} (\bibinfo{year}{1983}) \bibinfo{pages}{817--824}.
  \DOIprefix\doi{10.1002/spe.4380130908}.
\bibitem[{Johnson and Beekman(1988)}]{johnson1988}
\bibinfo{author}{M.~Johnson}, \bibinfo{author}{G.~Beekman},
  \bibinfo{title}{{Oh! Thinks Lightspeed Pascal!}}, \bibinfo{publisher}{WW
  Norton}, \bibinfo{year}{1988}.
\bibitem[{Ritchie et~al.(1988)Ritchie, Kernighan, and Lesk}]{ritchie1988}
\bibinfo{author}{D.~M. Ritchie}, \bibinfo{author}{B.~W. Kernighan},
  \bibinfo{author}{M.~E. Lesk}, \bibinfo{title}{{The C Programming Language}},
  \bibinfo{publisher}{Prentice Hall Englewood Cliffs}, \bibinfo{year}{1988}.
\bibitem[{Brooks(1978)}]{brooks1978}
\bibinfo{author}{R.~Brooks},
\newblock \bibinfo{title}{{Using a Behavioral Theory of Program Comprehension
  in Software Engineering}},
\newblock in: \bibinfo{booktitle}{Proceedings of the 3rd International
  Conference on Software Engineering (ICSE '78)}, \bibinfo{publisher}{IEEE
  Press}, \bibinfo{year}{1978}, p. \bibinfo{pages}{196–201}.
\bibitem[{Chance et~al.(1993)Chance, Zhuang, UnAh, Alter, and
  Lipton}]{chance1993}
\bibinfo{author}{B.~Chance}, \bibinfo{author}{Z.~Zhuang},
  \bibinfo{author}{C.~UnAh}, \bibinfo{author}{C.~Alter},
  \bibinfo{author}{L.~Lipton},
\newblock \bibinfo{title}{Cognition-activated low-frequency modulation of light
  absorption in human brain},
\newblock \bibinfo{journal}{Proceedings of the National Academy of Sciences}
  \bibinfo{volume}{90} (\bibinfo{year}{1993}) \bibinfo{pages}{3770--3774}.
  \DOIprefix\doi{10.1073/pnas.90.8.3770}.
\bibitem[{Peterson(1977)}]{peterson1977}
\bibinfo{author}{J.~L. Peterson},
\newblock \bibinfo{title}{{On the Formatting of Pascal Programs}},
\newblock \bibinfo{journal}{ACM SIGPLAN Notices} \bibinfo{volume}{12}
  (\bibinfo{year}{1977}) \bibinfo{pages}{83–86}.
  \DOIprefix\doi{10.1145/954618.954624}.
\bibitem[{Crider(1978)}]{crider1978}
\bibinfo{author}{J.~E. Crider},
\newblock \bibinfo{title}{{Structured Formatting of Pascal Programs}},
\newblock \bibinfo{journal}{ACM SIGPLAN Notices} \bibinfo{volume}{13}
  (\bibinfo{year}{1978}) \bibinfo{pages}{15–22}.
  \DOIprefix\doi{10.1145/953777.953779}.
\bibitem[{Gustafson(1979)}]{gustafson1979}
\bibinfo{author}{G.~G. Gustafson},
\newblock \bibinfo{title}{{Some Practical Experiences Formatting Pascal
  Programs}},
\newblock \bibinfo{journal}{ACM SIGPLAN Notices} \bibinfo{volume}{14}
  (\bibinfo{year}{1979}) \bibinfo{pages}{42–49}.
  \DOIprefix\doi{10.1145/988113.988118}.
\bibitem[{Ellis(2010)}]{Ellis2010}
\bibinfo{author}{P.~D. Ellis}, \bibinfo{title}{The Essential Guide to Effect
  Sizes: Statistical Power, Meta-Analysis, and the Interpretation of Research
  Results}, \bibinfo{publisher}{Cambridge University Press},
  \bibinfo{year}{2010}.
\bibitem[{Cohen(1992)}]{cohen1992}
\bibinfo{author}{J.~Cohen},
\newblock \bibinfo{title}{{Statistical Power Analysis}},
\newblock \bibinfo{journal}{Current Directions in Psychological Science}
  \bibinfo{volume}{1} (\bibinfo{year}{1992}) \bibinfo{pages}{98--101}.
  \DOIprefix\doi{10.1111/1467-8721.ep10768783}.
\bibitem[{Gopstein et~al.(2018)Gopstein, Zhou, Frankl, and
  Cappos}]{Gopstein2018}
\bibinfo{author}{D.~Gopstein}, \bibinfo{author}{H.~H. Zhou},
  \bibinfo{author}{P.~Frankl}, \bibinfo{author}{J.~Cappos},
\newblock \bibinfo{title}{{Prevalence of Confusing Code in Software Projects:
  Atoms of Confusion in the Wild}},
\newblock in: \bibinfo{booktitle}{Proceedings of the 15th International
  Conference on Mining Software Repositories (MSR '18)},
  \bibinfo{publisher}{ACM}, \bibinfo{address}{New York, NY, USA},
  \bibinfo{year}{2018}, pp. \bibinfo{pages}{281--291}.
  \DOIprefix\doi{10.1145/3196398.3196432}.

\end{thebibliography}
